\documentclass[11pt]{article}
\usepackage{epic}
\usepackage{curves}
\catcode`\@=11
\def\marginnote#1{}

\newcount\hour
\newcount\minute
\newtoks\amorpm
\hour=\time\divide\hour by60
\minute=\time{\multiply\hour by60 \global\advance\minute by-\hour}
\edef\standardtime{{\ifnum\hour<12 \global\amorpm={am}%
        \else\global\amorpm={pm}\advance\hour by-12 \fi
        \ifnum\hour=0 \hour=12 \fi
        \number\hour:\ifnum\minute<10 0\fi\number\minute\the\amorpm}}
\edef\militarytime{\number\hour:\ifnum\minute<10 0\fi\number\minute}

\def\draftlabel#1{{\@bsphack\if@filesw {\let\thepage\relax
      \xdef\@gtempa{\write\@auxout{\string
          \newlabel{#1}{{\@currentlabel}{\thepage}}}}}\@gtempa \if@nobreak
    \ifvmode\nobreak\fi\fi\fi\@esphack} \gdef\@eqnlabel{#1}}
    \def\@eqnlabel{}
\def\@vacuum{}
\def\draftmarginnote#1{\marginpar{\raggedright\scriptsize\tt#1}}

\def\draft{
  \oddsidemargin -.5truein
  \def\@oddfoot{\footnotesize \sl preliminary draft \hfil
    \rm\thepage\hfil\sl\today\quad\militarytime}
  \let\@evenfoot\@oddfoot \overfullrule 3pt
    \let\label=\draftlabel
    \let\marginnote=\draftmarginnote
  \def\@eqnnum{(\theequation)\rlap{\kern\marginparsep\tt\@eqnlabel}%
    \global\let\@eqnlabel\@vacuum}

  }
\hoffset= - 1.0in

\def\be{\begin{equation}}
\def\ee{\end{equation}}
\def\bea{\begin{eqnarray}}
\def\eea{\end{eqnarray}}
\def\<{\langle}
\def\>{\rangle}

\def\res{{{\rm res}}}
\def\Im{{\rm Im}}
\def\F{{\cal F}}

\def\d{\partial}
\def\N2{${\cal N}=2$}

\def\tr{{\mathrm{tr\,}}}

\def\1N{${\cal N}=1$}
\def\4N{${\cal N}=4$}

\def\e{{\,\rm e}\,}

\def\bea{\begin{eqnarray}}
\def\eea{\end{eqnarray}}
\def\bqa{\begin{eqnarray}}
\def\eqa{\end{eqnarray}}

\def\beq{\begin{equation}}
\def\eeq{\end{equation}}
\def\ba{\beq\begin{array}{c}}
\def\ea{\end{array}\eeq}
\def\be{\beq}
\def\ee{\eeq}

\let\text=\mathrm

\newcommand\theTag[1]{(\ref{#1})}
\def\e{e}

\def\beq{\begin{equation}}
\def\eeq{\end{equation}}
\def\bea{\begin{eqnarray}}
\def\eea{\end{eqnarray}}

\newcommand{\rf}[1]{(\ref{#1})}

\def\F {{\cal F}}

\def\eps{\varepsilon}

\renewcommand{\d}{{{\partial}}}

\renewcommand{\<}{\langle}
\renewcommand{\>}{\rangle}

\def\2{{1\over 2}}

\def\d{\partial}

\def\â{$\tau$}

\newcommand{\cpict}[3]{
\dimen1=#1\advance\dimen1 by-\hsize\divide\dimen1 by-2
\vtop to #2{
\noindent\hskip\dimen1{\special{em:graph #3.bmp}}
\vfil}\hskip-2cm
}

\newcommand{\dV}{\frac{\partial}{\partial V(\lambda)}}

\newcommand{\ty}{{\tilde y}}
\newcommand{\why}{{\widehat y}}

\let\@@savethanks\thanks
\def\thanks#1{\gdef\thefootnote{\alph{footnote}}\@@savethanks{#1}}

\title{{\bf Matrix models with hard walls: Geometry and solutions}
\vspace{.5cm}}
\author{{\bf L. Chekhov}\thanks{E-mail: \ chekhov@mi.ras.ru}
\date{ } \\ {\small
{\it Steklov Mathematical Institute, Moscow, Russia}}
\\
{\small{\it Institute for Theoretical and Experimental Physics, Moscow, Russia}},
\\
{\small{\it Poncelet Laboratoire International Franco--Russe, Moscow, Russia}},
\\
{\small and {\it Department of Mathematics and Statistics, Concordia University, Montreal, Canada}}
\\
}

\begin{document}

\maketitle

\vspace{-10.5cm}

\begin{center}
\hfill ITEP/TH-27/05\\
%\hfill hep-th/0602013
\end{center}

\vspace{9.5cm}

\begin{abstract}
We discuss various aspects of most general multisupport solutions to
matrix models in the presence of hard walls, i.e., in the case where
the eigenvalue support is confined to subdomains of the real axis.
The structure of the solution at the leading order
is described
by semiclassical, or generalized Whitham--Krichever hierarchies as in the unrestricted case.
Derivatives of tau-functions for these
solutions are associated with families of Riemann surfaces (with
possible double points) and satisfy the
Witten--Dijkgraaf--Verlinde--Verlinde equations.
We then develop the
diagrammatic technique for finding free energy of this model in all orders of the 't~Hooft
expansion in the reciprocal matrix size generalizing the Feynman diagrammatic technique
for the Hermitian one-matrix model due to Eynard.
\end{abstract}
%\tableofcontents
\def\thefootnote{\arabic{footnote}}

The matrix models and their so-called
multisupport (multicut) solutions became important again after the studies in
${\cal N}=1$ SUSY gauge theories \cite{CIV1}, \cite{CIV2}
and due to Dijkgraaf and Vafa \cite{DV}, who proposed to calculate the
low energy superpotentials using the partition function of multicut solutions.
These solutions, known since early 90th
(see, e.g., \cite{JU90,AkAm}), were revived by  Bonnet, David and Eynard
\cite{David}.

Dijkgraaf and Vafa proposed to consider the nonperturbative
superpotentials of ${\cal N}=1$ SUSY gauge theories in
four dimensions arising from the partition
functions of the one-matrix model (1MM) in the leading order in
$1/N$, $N$ being the matrix size. The leading order (of the
't~Hooft $1/N$-expansion) of the matrix model is described by the
semiclassical tau-function of the so-called universal Whitham--Krichever (WK)
hierarchy \cite{Kri1} (see also \cite{MarMir98,Mbook,SWbook}; the
details about one-matrix and two-matrix cases see in \cite{ChM} and
\cite{KM}). At the same time, matrix integrals beyond large-$N$ limit
are tau-functions of the
hierarchies of integrable equations of the KP/Toda type
\cite{IntMamo}. For the planar single-cut solutions, the matrix-model
partition functions become $\tau$-functions of the dispersionless
Toda hierarchy, one of the simplest example of the WK hierarchy.

One may also consider more general solutions to matrix models, identifying
them with generic solutions to the loop (Schwinger--Dyson, or Virasoro)
equations \cite{48}, or the Ward identities satisfied by matrix integrals
\cite{Virasoro}. We discuss an interesting class of
multi-cut, or multi-support, solutions to the loop equations possessing
multi-matrix integral representations
\cite{David,KMT,AMM,AMM3}. These solutions are associated with families
of Riemann surfaces and form a sort of a basis in the space of all
solutions to the loop equations \cite{AMM,AMM3} (like the finite-gap
solutions form a similar basis in the space of all solutions to an
integrable hierarchy). They can be distinguished by their
``isomonodromic" properties---switching on higher matrix model
couplings or $1/N$-corrections does not change the family of
Riemann surfaces, but just reparameterizes the moduli as functions
of these couplings. This property is directly related to that
the partition functions of these solutions are semiclassical
tau-functions (also called the prepotentials of the corresponding
Seiberg--Witten-like systems).

The geometrical properties of these solutions
were extensively studied in \cite{ChMMV,ChMMV2}, and the comprehensive procedure
for calculating $1/N$-expansion terms was developed in \cite{Ey},
\cite{CEy}. In this paper, we give the review of methods of these papers and
extend them to the case of matrix models with hard walls (hard edges). Those are
models in which we introduce rigid restrictions on the domain of admissible
eigenvalues of the potential. It is well-known that the behavior of the eigenvalue
distribution changes dramatically in the presence of such walls: it has inverse
square-root singularity as approaching the wall (see, e.g., Szeg{\o} \cite{Sz}).

Our aim in this paper is to show how the solutions with hard walls can be nicely
incorporated in the general setting of solutions to 1MM. Actually we show that all
the basic ingredients of 1MM solutions, that is, the WK hierarchy, the WDVV equations, and the
diagrammatic technique for evaluating corrections in $1/N$, turn out to be common for solutions with
and without hard walls and only the hyperelliptic
Riemann surface representing the spectral curve is changed. Technical solutions
however apparently develop distinctions, which we indicate in what follows.

Recently, models with hard walls in the planar limit of two-matrix models were
constructed~\cite{Ey1,Bertola}.

For the generality of presentation, we use the term 1MM to denote a Hermitian one-matrix model
possibly with hard walls.

In Sec.~\ref{s:1MM}, we describe the general properties of multi-cut solutions of
matrix models. We derive the basic set of constraints on the matrix model partition
functions and present them in the form of the master loop equation.

Before presenting the general procedure of solving the master loop equation order by order
to find the loop mean (one-point resolvents) of the matrix model potential, we present the
special investigation of just the leading (planar) approximation.
We prove that the free energy of
the 1MM in the planar (large-$N$) limit coincides with the
prepotential of a Seiberg--Witten--Whitham--Krichever theory.

The corresponding WK hierarchy is formulated in terms of Abelian differentials on
(a family of) Riemann surfaces. This implies that the main quantities in matrix
models can be expressed in geometrical terms and allow
calculating derivatives of the matrix model free energy. Indeed, we demonstrate
in Sec.~\ref{s:geometry} that the
second derivatives of the 1MM free energy
can be expressed through the so-called Bergmann bi-differential.
On this language we derive the general relation satisfied by the matrix model
potentials when differentiating w.r.t. Whitham times.
We then come to third derivatives.
Indeed, it is typical (but not compulsory) for WK
tau-functions to satisfy
a set of (generalized) Witten--Dijkgraaf--Verlinde--Verlinde
(WDVV) equations \cite{Wit90,LGMT,wdvvg} that are differential equations
with respect to Whitham times involving third derivatives.
These equations are usually considered an evidence for an
underlying topological theory.
In Sec.~\ref{s:geometry}, we prove
that the semiclassical tau-function of the multi-support solutions
to matrix models
satisfies the WDVV equations in the case of general 1MM solution, i.e., in
the case of arbitrary number of nonzero canonical times,
which include the times of the potential and the occupation numbers
(filling fractions), which
indicate the portions of eigenvalues that
dwell on the related intervals of the eigenvalue support.
The hard-wall parameters, although being out of the set of canonical variables,
affect nevertheless the {\em dimension} of the set of nondegenerate WDVV equations.
This completes an interpretation of the results of
\cite{CIV1,CIV2} in terms of semiclassical hierarchies.
The WDVV equations follow from
the residue formula and associativity of some algebra (e.g., of the
holomorphic differentials on the Riemann surface) \cite{wdvvg,wdvv2}.
We present the proof of the residue formula and extra conditions following
\cite{ChMMV}. Analogous results hold true in the two-matrix model~\cite{Bertola2}

In Sec.~\ref{s:genera}, we describe the diagrammatic procedure that allows
evaluating order by order all the corrections in $1/N$ in the 1MM with hard walls.
In the same section we describe the procedure of \cite{CEy} to integrate the
obtained answer to obtain the expression for the free energy itself. This procedure works in
all genera except the subleading (torus) approximation, which, correspondingly, deserves the special
investigation (as in the case of the standard 1MM). We therefore devote the
special section~\ref{s:genus} to the calculation of the free-energy subleading correction.
Note again that analogous formulas were obtained in the two-matrix model case by
Eynard, Kokotov, and Korotkin~\cite{EKK}.

\section{Matrix models and Riemann surfaces \label{s:1MM}}

\subsection{Matrix integrals and resolvents \label{ss:resolvent}}

Consider the 1MM integral
\be
%\int_{t_0 N\times t_0 N}DX\, \e^{-N\tr V(X)}=\e^{\cal F},
\int_{N\times N;\,E}DX\, \e^{-{1\over \hbar}\tr V(X)}=\e^{{1\over \hbar^2}\cal F}\equiv{\cal Z},
\label{Xap2.1}
\ee
%M->{1\over \hbar}, N= dim(X) (DV)
where $V(X)=\sum_{k\geq 1}^{}t_kX^k$, $\hbar = {t_0\over N}$ is
a formal expansion parameter, the integration goes
over the $N\times N$ matrices,
 $DX\propto\prod_{i<j}d\Re X_{ij}d\Im X_{ij}\prod_id\Re X_{ii}$, and
we also assume that the integration domain $E$ for the eigenvalues of the
matrix $X$ comprises a number of (possibly half-infinite) intervals,
$E=\cup_{\beta=1}^{q/2}[a_{2\beta-1},a_{2\beta}]$.
The topological expansion of the Feynman diagrams series is then equivalent to
the expansion in even powers of $\hbar$ for
\be
{\cal F}\equiv {\cal F}(\hbar,t_0, t_1, t_2, \dots)
=\sum_{h=0}^{\infty}{\hbar}^{2h}{\cal F}_h,
\label{Xap2.2}
\ee
Customarily $t_0=\hbar N$ is the scaled
number of eigenvalues. We assume the potential $V(\lambda)$ to be a polynomial
of the fixed degree $m+1$, with the fixed constant "highest" time $t_{m+1}=1$.

The averages, corresponding to the partition function~\theTag{Xap2.1} are
defined as usual:
\beq
\bigl\langle f(X)\bigr\rangle=
\frac1{\cal Z}\int_{N\times N;\,E}DX\,f(X)\,\exp\left(-{1\over \hbar}\tr V(X)\right)
\label{4.1}
\eeq
and it is convenient to use their
generating functionals: the one-point resolvent
\beq
W(\lambda)=%\frac{1}{M}
\hbar
\sum_{k=0}^{\infty}
\frac{\langle\tr X^{k}\rangle}{\lambda^{k+1}}
\label{4.2}
\eeq
as well as the $s$-point resolvents $(s\geq2)$
\beq
W(\lambda_1,\dots,\lambda_s)=
%M^{s-2}
\hbar^{2-s}
\sum_{k_1,\dots,k_s=1}^{\infty}
\frac{\langle\tr X^{k_1}\cdots\tr X^{k_s}\rangle_{\mathrm{conn}}}
{\lambda_1^{k_1+1}\cdots \lambda_s^{k_s+1}}=
%M^{s-2}
\hbar^{2-s}
\left\langle\tr\frac{1}{\lambda_1-X}\cdots
\tr\frac{1}{\lambda_s-X}\right\rangle_{\mathrm{conn}}
\label{4.3}
\eeq
where the subscript ``$\mathrm{conn}$" pertains to the connected
part.

These resolvents are obtained from the free energy ${\cal F}$ through the
action
\bea
W(\lambda_1,\dots,\lambda_s)&=&\hbar^2\frac{\d}{\d V(\lambda_s)}\frac{\partial}{\partial V(\lambda_{s-1})}\cdots
\frac{\partial {\cal F}}{\partial V(\lambda_1)}=\nonumber
\\
&=&\frac{\partial }{\partial V(\lambda_s)}\frac{\partial }{\partial V(\lambda_{s-1})}\cdots
\frac{\partial }{\partial V(\lambda_2)}W(\lambda_1),
\label{4.5}
\eea
of the loop insertion operator
\beq
\frac{\partial }{\partial V(\lambda)}\equiv
-\sum_{j=1}^{\infty}\frac{1}{\lambda^{j+1}}\frac{\d}{\d t_{j}}.
\label{4.6}
\eeq
Therefore, if one knows exactly the one-point resolvent for arbitrary
potential, all multi-point resolvents can be calculated by induction.
In the above normalization, the genus expansion has the form
\beq
W(\lambda_1,\dots,\lambda_s)=\sum_{h=0}^{\infty}
%\frac{1}{M^{2g}}
\hbar^{2h}
W_{h}(\lambda_1,\dots,\lambda_s),\quad s\geq1,
\label{4.7}
\eeq
which is analogous to genus expansion \rf{Xap2.2}.

\subsection{Master loop equation \label{ss:loopequation}}

We begin with deriving conditions on integral (\ref{Xap2.1}) coming from its
invariance w.r.t. changing the integration variables. For this, we first
to the eigenvalue representation assuming
\be
e^{\frac1{\hbar^2}\cal F}=\int_{E(\{a_\beta\})} DX e^{-\frac1\hbar\tr V(X)}\simeq
\int_{E(\{a_\beta\})}\prod_{i=1}^{t_0/\hbar} d\lambda_i\Delta^2(\lambda)
e^{-\frac1\hbar\sum_{i=1}^{t_0/\hbar}V(\lambda_i)}
\label{eig}
\ee
up to the volume of the unitary group $U(t_0/\hbar)$. Here, as usual, we let $\Delta(\lambda)$
denote the Vandermonde determinant of variables $\lambda_i$.
In the latter expression, we
perform the change of variables
\be
\lambda_i=\tilde\lambda_i+\eps\tilde\lambda_i^{p+1},\quad p\ge-1,
\label{eig*1}
\ee
obtaining
$$
e^{\frac1{\hbar^2}\cal F}=\int_{E(\{a_\beta-\eps a_\beta^{p+1}\})}
\prod_{i=1}^{t_0/\hbar} d\bigr(\tilde\lambda_i+\eps\tilde\lambda_i^{p+1}\bigl)
\Delta^2(\tilde\lambda+\eps\tilde\lambda^{p+1})
e^{-\frac1\hbar\sum_{i=1}^{t_0/\hbar}V\bigr(\tilde\lambda_i+\eps\tilde\lambda_i^{p+1}\bigl)}.
$$
Segregating the coefficient by $\eps$ in a standard way and equating it to zero, we obtain
the Virasoro conditions to be satisfied by integral (\ref{Xap2.1}) (see \cite{Adler})
\be
\left(-\sum_{\beta=1}^q a_\beta^{p+1}\frac{\d}{\d a_\beta}
+\sum_{k=0}^p\hbar^2\frac{\d}{\d t_k}\cdot\frac{\d}{\d t_{p-k}}+\sum_{k=1}^{\infty}kt_k
\frac{\d}{\d t_{k+p}}\right){\cal Z}=0,\quad p\ge-1,
\label{cond1}
\ee
where by definition
$$
\frac{\d}{\d t_0}{\cal Z}\equiv -\frac{t_0}{\hbar^2}{\cal Z},\hbox{ or }
\frac{\d}{\d t_0}{\cal F}\equiv -t_0.
$$

In terms of ${\cal F}$, Eqs. (\ref{cond1}) become
\be
-\sum_{\beta=1}^q a_\beta^{p+1}\frac{\d{\cal F}}{\d a_\beta}
+\sum_{k=0}^p\frac{\d{\cal F}}{\d t_k}\frac{\d{\cal F}}{\d t_{p-k}}
+\hbar^2\sum_{k=0}^p\frac{\d^2{\cal F}}{\d t_k\d t_{p-k}}
+\sum_{k=1}^{\infty}kt_k
\frac{\d{\cal F}}{\d t_{k+p}}=0,\quad p\ge-1,
\label{cond2}
\ee
and contracting these {\em Virasoro conditions} with $\lambda^{-p-2}$, we obtain
the master loop equations \cite{48} of the 1MM with hard walls~\cite{Virasoro}:
\beq
\sum_{\beta=1}^q \frac1{\lambda-a_\beta}\frac{\d{\cal F}}{\d a_\beta}
+\oint_{{\cal C}_{\cal D}}\frac{d\xi}{2\pi
i}\frac{V'(\xi)}{\lambda-\xi}W(\xi)=
W(\lambda)^2+\hbar^2 W(\lambda,\lambda).
\label{4.8}
\eeq
We introduce the linear integral operator $\widehat{K}$,
\beq
\widehat{K}f(\lambda)\equiv\oint_{{\cal C}_{\cal D}}\frac{d\xi}{2\pi i}
\frac{V'(\xi)}{\lambda-\xi}f(\xi)=\left[V'(\lambda)f(\lambda)\right]_-
\label{4.12}
\eeq
projects onto the negative powers\footnote{In order to prove it, one
suffices to deform the integration contour to infinity to obtain
$$
\oint_{{\cal C}_{\cal D}}\frac{d\xi}{2\pi i}
\frac{V'(\xi)}{\lambda-\xi}f(\xi)=V'(\lambda)f(\lambda)-\left[V'(\lambda)f(\lambda)\right]_+=
\left[V'(\lambda)f(\lambda)\right]_-.
$$
}
of $\lambda$.
Hereafter, ${\cal C}_{\cal D}$~is a contour encircling all singular points
of $W(\xi)$, but not the point
$\xi=p$.
%; this contour integration acts as the projection operator
%extracting the coefficient
%of the term~$p^{-1}$.
Using Eq.~\theTag{4.5}, one can express
the second term in the r.h.s.\ of loop equation~\theTag{4.8} through
$W(\lambda)$, and Eq.~\theTag{4.8} becomes then an equation for
one-point resolvent \rf{4.2}.

Substituting genus expansion~\theTag{4.7} in Eq.~\theTag{4.8}, one finds
that $W_h(\lambda)$ for $h\geq1$ satisfy the equation
\beq
\sum_{\beta=1}^q \frac1{\lambda-a_\beta}\frac{\d{\cal F}_h}{\d a_\beta}
+\left(\widehat{K}-2W_{0}(\lambda)\right)W_h(\lambda)=\sum_{h'=1}^{h-1}
W_{h'}(\lambda)W_{h-h'}(\lambda)+\frac{\partial }{\partial V(\lambda)}W_{h-1}(\lambda),
\label{4.11}
\eeq
In Eq.~\theTag{4.11}, $W_h(\lambda)$ is expressed through only
$W_{h_i}(\lambda)$ for which $h_i<h$. This fact allows
developing the iterative procedure.

Note that the loop equation contains now explicit dependence on the points $a_\beta$. Although
the term proportional to the derivative of ${\cal F}_h$ seems not to be universal, in what
follows we find that these terms are important to keep the picture self-consistent.

We use the asymptotic condition
(which follows from the definition of the matrix
integral)
\be
W_h(\lambda)|_{\lambda\to\infty} = \frac{t_0}{\lambda}\delta_{h,0}+O({1}/{\lambda^2}),
\label{Winf}
\ee
and manifestly solve (\ref{4.8}) for genus zero. Then, one could
iteratively find $W_h(\lambda)$ thus restoring the corresponding contributions
in the free energy by integration, since
\be
\label{total}
W_h(\lambda)=-\dV {\cal F}_h,\quad h\ge 1.
\ee

\subsection{Solution in genus zero} \label{basic}

In genus zero, loop equation (\ref{4.8}) becomes
\be
\sum_{\beta=1}^q \frac1{\lambda-a_\beta}\frac{\d{\cal F}_0}{\d a_\beta}
+\oint_{{\cal C}_{\cal D}}\frac{d\xi}{2\pi i}
\frac{V^{\prime}(\xi)}{\lambda-\xi} W_0(\xi)=
\left[V'(\lambda)W_0(\lambda)\right]_-
+\frac{{\overline P}_{q-1}(\lambda)}{\prod_{\beta=1}^{q}(\lambda-a_\beta)}
= (W_0(\lambda))^2,
\label{plan}
\ee
where the polynomial ${\overline P}_{q-1}(\lambda)$ is such that
\be
\sum_{\beta=1}^q \frac1{\lambda-a_\beta}\frac{\d{\cal F}_0}{\d a_\beta}
=\frac{{\overline P}_{q-1}(\lambda)}{\prod_{\beta=1}^{q}(\lambda-a_\beta)}.
\label{polF}
\ee

In what follows, for simplicity, we let $P_r(\lambda)$ denote {\em any} polynomial of $\lambda$
of order $r$ (which is indicated by the subscript). For instance, we can write $P_r(\lambda)+
P_u(\lambda)=P_w(\lambda)$, where $w=\max(r,u)$.

In order to solve Eq. (\ref{plan}), note that
\be
\left[V'(\lambda)W_0(\lambda)\right]_-=V'(\lambda)W_0(\lambda)-\left[V'(\lambda)W_0(\lambda)\right]_+.
\ee
Due to (\ref{Winf}), the last term in the r.h.s. is a polynomial
of degree $m-1$, $m$ being the degree of $V'(\lambda)$,
\be
\label{polP}
P_{m-1}(\lambda)  = -\left[V'(\lambda)W_0(\lambda)\right]_+=
-\oint_{\cal C_{\infty}}\frac{d\xi}{2\pi i}
      \frac{V'(\xi)}{\lambda-\xi}W_0(\xi).
\ee
That is, we have
$$
W_0^2(\lambda)-V'(\lambda)W_0(\lambda)=P_{m-1}(\lambda)+
\frac{{\overline P}_{q-1}(\lambda)}{\prod_{\beta=1}^{q}(\lambda-a_\beta)}
=\frac{P_{m+q-1}(\lambda)}{\prod_{\beta=1}^{q}(\lambda-a_\beta)}.
$$

Then, the solution to \rf{plan} is
\be
W_0(\lambda) = \frac{1}{2}V'(\lambda) - \frac{1}{2}\frac{\sqrt{V'(\lambda)^2\prod_{\beta=1}^{t}(\lambda-a_\beta)
+4P_{m+t-1}(\lambda)}}{\sqrt{\prod_{\beta=1}^{t}(\lambda-a_\beta)}},
 \label{*loop2*}
\ee
where the minus sign is chosen in order to fulfill asymptotics (\ref{Winf}).
We also assume that the general solution in the large-$N$ limit
depends only on $t$ among $q$ hard wall points $a_\beta$. Then, from (\ref{polF}),
we have that the corresponding pole disappear and does not enter expression (\ref{*loop2*}).
We therefore present the one-point resolvent in the form
\be
W_0(\lambda) = \frac{1}{2}\left( V'(\lambda)-y(\lambda)\right),
\label{*loop3*}
\ee
therefore introducing a new function $y(\lambda)$ satisfying the equation
\be
y(\lambda)^2=V'(\lambda)^2+4\frac{P_{m+t-1}(\lambda)}{\prod_{\beta=1}^{t}(\lambda-a_\beta)}
\label{1mamocu}
\ee
and determined on a hyperelliptic Riemann
curve. For the generic potential $V(\lambda)$ with $m\to\infty$, this curve
may have an infinite
genus, but we are going to consider solutions with only finite number $n$ of cuts.
Endpoints of these cuts are $t$ points $a_\beta$ \ $(t\ge q)$, which we enumerate
by $\beta=1,\dots,t$, and $s$ new endpoints $\mu_\alpha$, \ $\alpha=1,\dots,s$.
Apparently, $s+t=2n$. We then present $y(\lambda)$ in the form
\be
\label{ty}
y(\lambda)\equiv M(\lambda)\ty(\lambda),
\quad \hbox{and } \quad
{\ty}^2(\lambda)\equiv\frac{\prod_{\alpha=1}^{s}(\lambda-\mu_\alpha)}
{\prod_{\beta=1}^{t}(\lambda-a_\beta)}
\ee
with all $\mu_\alpha$, $a_\beta$ distinct.
In what follows, we assume
$M(\lambda)$ to be a polynomial of degree $m-\frac{s-t}2$, keeping in mind that both $s$ and
$t$ are always
finite and fixed, while $m\geq n$ can be chosen arbitrarily large.
By convention, we set $\ty(\lambda)|_{\lambda\to\infty}\sim \lambda^{(s-t)/2}$, and
due to (\ref{*loop3*}),
\be\label{t3}
V'(\lambda)= y(\lambda) + {2t_0\over \lambda}+O(1/\lambda^2)
\ee
at large $\lambda$.

In comparison with the standard 1MM situation, we may encounter an ambiguity. Indeed, as $\ty(\lambda)$
is proportional to $\lambda^{(s-t)/2}$ as $\lambda\to\infty$, multiplying $W_0(\xi)$ by
$\frac{\ty^{-1}(\xi)}{\xi-\lambda}$ and integrating at infinity may produce a nonzero result for $t>s$.
For $M(\lambda)$, we then have
\be
M(\lambda) = \oint_{\cal C_{\infty}} \frac{d\xi}{2\pi i}
\frac{V'(\xi)+P_{\frac{t-s}2}(\xi)-P_{\frac{t-s}2}(\lambda)}{(\xi-\lambda)\ty(\xi)}
\equiv \prod_i^{m-\frac{s-t}2}\left(\lambda-b_i\right).
\label{M}
\ee
Here $P_{\frac{t-s}2}(\xi)$ is a polynomial and $b_i$ are zeros of $M(\lambda)$. In our consideration,
we systematically avoid any explicit dependence on $b_i$. Instead, the relevant objects on which
the free energy depends are {\it moments\/} $M_{i}^{(p)}$ of the general matrix model potential
(see~\cite{ACKM,Ak96,KMT}) representing $(p-1)$st derivatives of this
polynomial at the branching points that are endpoints of intervals of eigenvalue distribution,
\be
\label{M2}
M_{i}^{(p)}=
\frac{1}{(p-1)!}\,\left.\left(\frac{\d^{p-1}}{\d\lambda^{p-1}}
M(\lambda)\right)\right|_{\lambda=\mu_i}, \ \hbox{e.g.,}\
M_{i}^{(1)}=M(\mu_i).
\ee
Among endpoints $\mu_i$ of these intervals there are hard wall points $a_\beta$, \
$\beta=1,\dots,t$ $(t\le q)$ and a number of new ``dynamical'' endpoints $\mu_\alpha$,
$\alpha=1,\dots,s$ with $s+t=2n$. Often (but {\em not} always) these points can be
considered on the equal footing, as in formula (\ref{M2}), and we reserve the
unified notation $\mu_i$ with the Latin subscript $i=1,\dots,2n$ to indicate any
branching point, whatever origin it has.

Directly from definition (\ref{M2}) follows the representation for the moments
\be
\label{M1}
M_{i}^{(p)}=\oint_{{\cal C}_{\cal D}}\frac{d\xi}{2\pi i}
\frac{V'(\xi)+P_{\frac{t-s}2}(\xi)-P_{\frac{t-s}2}(\lambda)}
{(\xi-\mu_i)^p\tilde y(\xi)},\quad p\ge1.
\ee

Inserting solution (\ref{ty}), (\ref{M}) into (\ref{*loop3*}) and deforming
the contour, we obtain the planar one-point resolvent
with an $n$-cut structure,
\be
W_0(\lambda) = \frac{1}{2}\oint_{{\cal C}_{\cal D}}\frac{d\xi}{2\pi i}
\frac{V'(\xi)+P_{\frac{t-s}2}(\xi)-P_{\frac{t-s}2}(\lambda)}
{\lambda-\xi}\frac{\ty(\lambda)}{\ty(\xi)},
\quad \lambda\not\in{\cal D}.
\label{W0}
\ee
The contour ${\cal C_{\cal D}}$ of integration here
encircles the finite number~$n$ of disjoint intervals
\be
{\cal D} \equiv \bigcup_{i=1}^n [\mu_{2i-1},\mu_{2i}],
\quad \mu_1< \mu_2< \ldots < \mu_{2n}.
\ee

We now address the question how many free parameters do we have.
In (\ref{W0}) we have $m$ times $t_k$ of the potential $V'_m(\xi)$
(recall that the highest time is the unity), $\frac{t-s}2-1$ coefficients
of the polynomial $P_{\frac{t-s}2}(\xi)$ (the constant term does not
contribute) and $s$ new endpoints $\mu_\alpha$. However, we must take into
account asymptotic conditions (\ref{Winf}), which impose the restrictions
\be
\oint_{{\cal C}_{\cal D}}\frac{d\xi}{2\pi i}
\frac{\xi^w V^{\prime}(\xi)}{\ty(\xi)} = 2t_0\delta_{\frac{s-t}2,w},
\quad w=0,1,\dots,\frac{s-t}2,
 \label{Rand1}
\ee
if $s\ge t$, or
\be
\oint_{{\cal C}_{\cal D}}\frac{d\xi}{2\pi i}
\frac{a_{\frac{t-s}2}}{\ty(\xi)} = 2t_0,
\quad \hbox{where}\quad P_{\frac{t-s}2}(\lambda)=a_{\frac{t-s}2}\lambda^{\frac{t-s}2}+\dots,
\label{Rand2}
\ee
for $s<t$. If we do not count the hard wall parameters,
in the first case, we have $m$ times of the potential, $s$ points $\mu_\alpha$, one
zero time $t_0$ minus $\frac{s-t}2+1$ restrictions, that is $m+\frac{s+t}2=m+n$
parameters. In the second case, we have $m$ times of the potential, $s$ points
$\mu_\alpha$, one zero time $t_0$, $\frac{t-s}2$ relevant coefficients of the
polynomial $P_{\frac{t-s}2}(\lambda)$ and one restriction, which gives again
$m+\frac{s+t}2=m+n$ free parameters. That means that, except the time $t_0$, we
have an $n-1$-dimensional space of parameters. We can arbitrary choose coordinates on this
space, but it turns out that there exists a distinguished set of $n-1$ independent variables that
parameterize solutions to the loop equations \cite{David,DV},
\be
\label{Sfr}
S_i =
\oint_{A_i}\frac{d\lambda}{4\pi i}\,y(\lambda)=
\oint_{A_i}\frac{d\lambda}{4\pi i}M(\lambda)\ty(\lambda),
\ee
where $A_i$, $i=1,\dots,n-1$ is the basis of $A$-cycles on the {\em reduced hyperelliptic
Riemann surface} of genus $h\equiv n-1$
\be
\label{tyred}
\why^2=\prod_{\alpha=1}^{s}(\lambda-\mu_\alpha)\prod_{\beta=1}^{t}(\lambda-a_\beta)\equiv
\prod_{i=1}^{2n}(\lambda-\mu_i).
\ee
(We may conveniently choose the $A$-cycles to be the first $n-1$ cuts, see Fig.1.)
Note that we do not distinguish the branching points $\mu_i$ in the above expression by
their origin: both ``dynamical'' branching points $\mu_\alpha$ and hard walls $a_\beta$
enter $\why(\lambda)$ on the equal footing.
in contrast to the standard 1MM case, $\why(\lambda)\ne \ty(\lambda)$. However, in what
follows, $\why(\lambda)$ enters some of the constructions of the paper.

Given the basis of $A$-cycles, we also choose the conjugated basis of
$B$-cycles with the intersection form $A_i\circ B_j=\delta_{ij}$.
Besides canonically conjugated $A$- and $B$-cycles, we also use the linear
combination of $B$-cycles: $\bar B_i\equiv B_i-B_{i+1}$, $\bar B_{n-1}\equiv
B_{n-1}$. Therefore, $\bar B$-cycles encircle the nearest ends of two
neighbor cuts, while all $B$-cycles goes from a given right end of the cut
to the last, $n$th cut. For the sake of definiteness, we order all points
$\mu_i$ in accordance with their index so that $\mu_i$ is to the right of
$\mu_j$ if $i>j$.

\begin{figure}[h]\label{fig9.1}
\begin{picture}(190,55)(10,10)
\multiput(40,40)(40,0){5}{\oval(30,10)}
\multiput(28,40)(40,0){5}{\line(1,0){24}}
\multiput(28,40)(40,0){5}{\circle*{2}}
\multiput(52,40)(40,0){5}{\circle*{2}}
%%B_1
\thinlines
\qbezier(47,40)(47,48)(60,48)
\qbezier(60,48)(73,48)(73,40)
\thicklines
\qbezier[10](47,40)(47,32)(60,32)
\qbezier[10](60,32)(73,32)(73,40)
%%B_2
\thinlines
\qbezier(87,40)(87,48)(100,48)
\qbezier(100,48)(113,48)(113,40)
\thicklines
\qbezier[10](87,40)(87,32)(100,32)
\qbezier[10](100,32)(113,32)(113,40)
%%B_3
\thinlines
\qbezier(127,40)(127,46)(136,47)
\qbezier(144,47)(153,46)(153,40)
\thicklines
\qbezier[7](127,40)(127,34)(136,33)
\qbezier[7](144,33)(153,34)(153,40)
%%B_4
\thinlines
\qbezier(167,40)(167,48)(180,48)
\qbezier(180,48)(193,48)(193,40)
\thicklines
\qbezier[10](167,40)(167,32)(180,32)
\qbezier[10](180,32)(193,32)(193,40)
\put(39,20){$A_1$}
\put(59,23){$\bar B_1$}
\put(39,50){$S_1$}
\put(79,20){$A_2$}
\put(99,23){$\bar B_2$}
\put(79,50){$S_2$}
\put(119,20){$A_3$}
\put(119,50){$S_3$}
\put(136,40){$\dots$}
\put(158,20){$A_g$}
\put(171,23){$\bar B_g{=}B_g$}
\put(159,50){$S_g$}
\put(199,20){$A_n$}
\put(189,50){$t_0-\sum S_i$}
% end of 1a
%
\end{picture}
\caption{Structure of cuts and contours for the reduced Riemann surface.}
\end{figure}
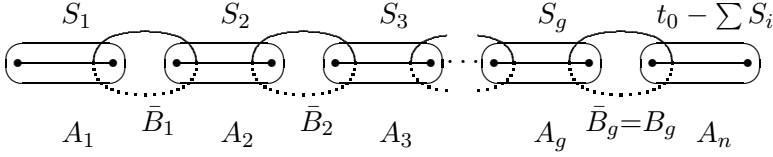

\subsection{Matrix eigenvalue picture}

The variables $S_i$ find a
``physical" interpretation in the semi-classical picture of matrix eigenvalues
(Coulomb gas) in the limit where their number (and, therefore, size of the matrix) goes to
infinity.

Let us first introduce the averaged eigenvalue distribution
\be\label{rho}
\rho(\lambda)\equiv {t_0\over N}\sum_i^N \left< \delta (\lambda-\lambda_i)\right>=
\frac{1}{2\pi i}\lim_{\epsilon\to 0}
         \Big( W(\lambda-i\epsilon)-W(\lambda+i\epsilon) \Big)
\ee
with $\lambda_i$ being eigenvalues of the matrix $X$. In the planar limit,
this quantity becomes
\be
\rho_0(\lambda)=\frac{1}{2\pi i}\lim_{\epsilon\to 0}
         \Big( W_0(\lambda-i\epsilon)-W_0(\lambda+i\epsilon) \Big)
= \frac{1}{2 \pi}\,\hbox{Im\,}y(\lambda)
\ee
and it satisfies the
equation\footnote{Indeed, by definition
$$
\oint_{\infty}{W_0(\lambda)\over x-\lambda}d\lambda=0.
$$
Now, using (\ref{*loop3*}) and definition (\ref{rho}) and pulling out
the contour from infinity, one easily comes to this equation.
}
\be\label{rho-eq}
\not\!\!\int_{\cal D}
 {\rho_0(\lambda)\over x-\lambda}d\lambda={1\over 2}V'(x),\ \ \ \ \forall
p\in {\cal D}.
\ee
This averaged eigenvalue distribution becomes the distribution of
eigenvalues in the limit where their number goes to infinity.
Then, matrix integral (\ref{eig}) becomes
\be
e^{\frac1{\hbar^2}{\cal F}}
=\int_{E}\prod_i dx_i e^{-{1\over \hbar^2}
\left(\int V(\lambda)\varrho(\lambda)
- \int \varrho (\lambda)\varrho(\lambda')\log |\lambda-\lambda'|
d\lambda d\lambda '\right)}\equiv \int_{E}\prod_i dx_ie^{{1\over
\hbar^2}S_{\mathrm{eff}}}
\label{varrho}
\ee
where we introduced the eigenvalue distribution
\be
\varrho (\lambda)\equiv {t_0\over N}\sum_i\delta(\lambda-\lambda_i).
\ee
Now, in the limit of large $N$, one can use the saddle point approximation
to obtain the equation for $\varrho (\lambda)$. Since
the variable $t_0$ plays the role of the (normalized) total number of eigenvalues,
\be
\label{t0}
t_0 = \frac1{4\pi i} \oint_{\cal C_{\cal D}}y(\lambda)d\lambda =
-\frac{1}2\res_\infty(yd\lambda)
\ee
and the support ${\cal D}$ of $\varrho$ consists of $n$ segments ${\cal D}_i$,
we impose the constraint
\be
\int\varrho(\lambda)d\lambda=t_0,
\ee
and, following \cite{David,DV}, fix the {\em occupation numbers} of eigenvalues in
each of the segments, $S_i$ (\ref{Sfr}), $i=1,\dots,n-1$. We assume the
occupation number for the last, $n$th cut to be $t_0-\sum_{i=1}^{n-1}S_i\equiv S_n$.
\footnote{It is sometimes convenient to consider $S_n$ instead of $t_0$ as a canonical variable.
However, in all instants we use $S_n$, we specially indicate it for not confusing $S_n$ with the ``genuine''
filling fraction variables $S_i$, $i=1,\dots,n-1$.}
(Obviously, no new parameters $S_i$ arise in the one-cut case.)
We formally attain this by introducing the corresponding {\em chemical potentials}
(Lagrange multipliers) $\Pi_0$ and $\Pi_i$, \
$i=1,\dots, n-1$, in the variational problem for the free energy,
which therefore becomes in the planar limit
\bea
{S}_{\mathrm{eff}} \left[\varrho;S_i,t_0,t_k\right] &=&
-\int_{\cal D} V(\lambda)\varrho(\lambda)d\lambda +
\int\!\!\int_{\cal D} \varrho(\lambda)
\log\left|\lambda-\lambda '\right|
\varrho (\lambda ')d\lambda d\lambda ' +
\nonumber
\\
&&-\Pi_0 \left(\int_{\cal D}\varrho(\lambda)d\lambda - t_0\right)
-\sum_{i=1}^{n-1} \Pi_i \left(\int_{D_i}\varrho(\lambda)d\lambda - S_i\right),
\label{variF}
\eea
while the saddle-point equation becomes
\be\label{varrho-eq'}
2\int \varrho(\lambda) \log|x-\lambda|d\lambda=V(x)+\Pi_i +\Pi_0,\ \ \ \ \forall
x\in {\cal D}_i,
\ee
and its derivative still coincides with (\ref{rho-eq}).

Therefore, with generic values of the constants $\Pi_i$,
$\varrho_c(\lambda)$ gives the general solution to (\ref{rho-eq}) (or the
planar limit of the loop equation): these constants describe
the freedom arising when solving the loop equation. However, in the matrix model
integral (where there are no any chemical potentials)
one would further vary ${\cal F}_0$ w.r.t. $\Pi_i$ to find the
``true" minimum of the eigenvalue configuration,
\be\label{oc}
{\d {\cal F}_0\over \d S_i}=0,\ \ \ \ \forall i.
\ee
This is a set of equation that fixes concrete values of $S_i$ and $\Pi_i$ in
the matrix integral.

Let us now calculate the derivative of ${\cal F}_0$ (\ref{variF}) w.r.t. $S_i$. From
(\ref{variF}), one has
\be
\left.\frac{\partial S_{\mathrm{eff}}}{\partial S_i}\right|_{\varrho=\rho_0}=
-\int_{\cal D}d\lambda \frac{\partial \rho_0(\lambda)}{\partial S_i}
\left(V(\lambda)-2\int_{\cal D}d\lambda'\log(\lambda-\lambda')\rho_0(\lambda')\right).
\label{*b2*}
\ee
The expression in the brackets on the r.h.s. of~(\ref{*b2*}) is almost
a variation of (\ref{variF}) w.r.t. the eigenvalue density, which is
\bea
0=\left.\frac{\delta S_{\mathrm{eff}}}{\delta \rho(\lambda)}\right|_{\varrho=\rho_0}&=& V(\lambda)-2\int_{\cal
D}d\lambda'\log(\lambda-\lambda')\rho_0(\lambda')+\Pi_i+\Pi_0 \nonumber
\\
&&\quad\hbox{for}\ \lambda\in D_i\subset{\cal D}.
\eea
It is therefore a step
function $h(\lambda)$, which is constant equal to $\zeta_i\equiv-\Pi_0-\Pi_i$
on each cut $A_i$. One then has
\bea
\frac{\partial {\cal F}_0}{\partial S_i}=
\left.\frac{\partial S_{\mathrm{eff}}[\varrho]}{\partial S_i}\right|_{\varrho=\rho_0}=-\int_{\cal
D}d\lambda
\frac{\partial \rho_0(\lambda)}{\partial S_i}h(\lambda)=
-\frac{1}{4\pi i}\sum_{j=1}^n\zeta_j{\partial\over\partial S_i}\oint_{A_j}y(\lambda)d\lambda=\nonumber\\
=-\sum_{j=1}^n\zeta_j{\partial S_j\over\partial S_i}=-\zeta_i+\zeta_n=\Pi_i.
\label{Dvdual}
\eea
In particular,
\be\label{Pi0}
{\d{\cal F}_0\over\d t_0}=\Pi_0
\ee

In \cite{JU90} it was proved that
the difference of values of $\Pi_i$
on two neighbor cuts is equal to
\footnote{The simplest way to prove it is to define function $h(\lambda)$ outside the cuts: $h(\lambda)=V(\lambda)-2\int_{\cal
D}d\lambda'\log({\lambda-\lambda'})\rho_0(\lambda')$ and note that $\left.h'(\lambda)\right|_{\lambda\notin {\cal D}}=2W_0(\lambda)$.}
\be
\label{xi}
\zeta_{i+1}-\zeta_i=2\int_{\mu_{2i}}^{\mu_{2i+1}}W_0(\lambda)d\lambda,
\ee
i.e.,
\bea
\Pi_i&=&(\zeta_{i+1}-\zeta_i)+(\zeta_{i+2}-\zeta_{i+1})+\ldots+(\zeta_{n-1}-\zeta_{n-2})+(\zeta_n-\zeta_{n-1})
\nonumber
\\
&=&\oint_{\bar B_i\cup \bar B_{i+1}\cup\dots\cup \bar B_g}y(\lambda)d\lambda
=\oint_{B_i}y(\lambda)d\lambda.
\label{Dvdual1}
\eea

The planar limit of the free energy can be obtained
by substituting the saddle point solution $\varrho$ into
(\ref{variF}):
\be\label{FS}
{\cal F}_0=S_{\mathrm{eff}}\left[\varrho_c\right]=
-{1\over 2}\int_{\cal D} V(\lambda)\varrho_c(\lambda)d\lambda
+{1\over 2}\Pi_0 t_0 +{1\over 2}\sum_{i=1}^{n-1} \Pi_i S_i.
\ee

In the paper, we choose the solution to the loop equation with fixed
occupation numbers, (\ref{Sfr}). Note that fixing the chemical potentials
(\ref{Dvdual})-(\ref{Dvdual1}) instead of (\ref{oc}), we
obtain the solution with the interchanged $A$- and $B$-cycles
on the Riemann surface (\ref{1mamocu}). However, ${\cal F}_0$ is {\em not} modular
invariant.
Under the change of homology basis, ${\cal F}_0$ transforms in accordance with the
duality transformations, see \cite{dWM}.
The higher-genus corrections become also basis-dependent: choosing
$S_i$ or $\Pi_i$ as independent variables, one obtains
different expressions, say, for the genus-one free energy, see Sec.~4.3.

Note that the presence of hard walls has had no effect on consideration in this section
because it only changes asymptotic behavior of the function $y(\lambda)$ near the corresponding
branching points: while for ``dynamical'' branching points $\mu_\alpha$ the eigenvalue
distribution undergoes the semicircular Wigner's law, it is known (see, e.g., \cite{Sz}) that
in the vicinity of hard wall we observe accumulation of eigenvalues (zeros of the
corresponding orthogonal polynomials) with the reciprocal square-root law. This perfectly matches
with our answer (\ref{ty}) for $y(\lambda)$.

In the next subsection, we demonstrate that the planar loop equation
solution with fixed occupation numbers corresponds to a
Seiberg--Witten--Whitham--Krichever system, just as in the case of ordinary 1MM \cite{ChMMV}.

\subsection{Free energy as the prepotential of SWWK system \label{ss:prepotential}}

We now turn to studying geometrical properties hidden in the matrix-model solutions. Namely, we
associate a Seiberg--Witten--Whitham--Krichever (SWWK) system with the planar limit of the matrix-model
free energy.

\paragraph{Matrix integral as a Seiberg--Witten system.}

The family of Riemann surfaces is now the family of $g=n-1$
reduced Riemann surfaces described by (\ref{1mamocu}) or (\ref{ty}).
This means that these Riemann surfaces contain no information about
the additional polynomial $M(\lambda)$, which
is present, however, through \rf{ty} in the differential $dS$.
The role of SW differential is played by
\be\label{dS}
dS=y(\lambda)d\lambda.
\ee
We now consider its variation w.r.t. $S_i$.
Variations over moduli of the surface
do not change the genus of the reduced Riemann surface as
well as the highest degree of the additional polynomial $M_{m-n}(\lambda)$.
We also consider {\em both} the times of the potential $V'_m(\lambda)$ {\em and} the hard
wall parameters $a_\beta$ to be {\it independent\/} on the parameters $S_i$ and $t_0$,
that is, we assume $\delta V'/\delta S_i\equiv0$ and consider the general
variation $\delta$, which varies the potential and the branching points $\mu_\alpha$
(but {\em not} the hard wall parameters $a_\beta$ and the special variation
$\delta_S$ that leaves invariant the potential and the hard wall parameters and
change only the moduli parameters $S_i$.

Using (\ref{1mamocu}), (\ref{ty}), one obtains for the
{\it general\/} variation~$\delta dS$:
\footnote{Note that the variation $\delta$ differs nevertheless from loop
insertion (\ref{4.6}) because
the former does not change, by definition, the degree of the
polynomial $M(\lambda)$.}
\bea
\delta dS&=&\delta\left( M_{m-\frac{s-t}2}(\lambda)
{\tilde y}(\lambda)\right)d\lambda=
\nonumber
\\
&=&{\prod_{\alpha=1}^{s}(\lambda-\mu_\alpha)
\delta M_{m-\frac{s-t}2}(\lambda)+
{1\over 2}M_{m-\frac{s-t}2}(\lambda)
\delta \prod_{\alpha=1}^{s}(\lambda-\mu_\alpha) \over {\why}(\lambda)}d\lambda,
\label{v1}
\eea
where the polynomial expression in the numerator is of maximum
degree $m+\frac{s+t}2-1=m+n-1$ (since the
highest term of $M_{m-\frac{s-t}2}$ is fixed) and we have in the
denominator the hyperelliptic coordinate (\ref{tyred}).
On the other hand, under~$\delta_S$ which
does not alter the potential, we obtain
from (\ref{1mamocu}), (\ref{ty}) that
\bea
\delta_S dS&=&-{1\over 2}{\delta_S 4P_{m+t-1}(\lambda)\over
M_{m-\frac{s-t}2}(\lambda){\tilde y}(\lambda)\prod_{\beta=1}^t(\lambda-a_\beta)}d\lambda
\nonumber
\\
&=&-2{\delta_S P_{m+t-1}(\lambda)\over
M_{m-\frac{s-t}2}(\lambda){\why}(\lambda)}d\lambda.
\label{v2}
\eea
Because this variation is just a particular case of (\ref{v1}), we obtain
that zeros of $M_{m-\frac{s-t}2}(\lambda)$ in the
denominator of (\ref{v2}) must be exactly {\it cancelled} by zeros of the
polynomial $\delta_S P_{m+t-1}(\lambda)$ in the numerator,
so the maximum degree of the polynomial in
the numerator is $\frac{s+t}2-2=n-2$ (because, again,
the highest-order term of $P_{m+t-1}(\lambda)$ is fixed by asymptotic
condition (\ref{t3}) and is not altered by variations $\delta_S$).
We then come to the crucial observation that the variation
$\delta_S dS$ is {\it holomorphic} on curve (\ref{tyred}), as it should be
for the SW differential.

Moreover, considering specific variation w.r.t. $S_i$ with condition (\ref{Sfr})
taken into account, we find that
$$
\delta_{i,j}=\delta_{S_i}S_j=
\delta_{S_i}\oint_{A_j}\frac{d\lambda}{4\pi i}y(\lambda)=
\oint_{A_j}\frac{d\lambda}{4\pi i}\delta_{S_i}y(\lambda)
$$
and hence
\be
{\d dS\over\d S_i}={ H_i(\lambda)d\lambda\over\why(\lambda)}=d\omega_i
\label{canoDV}
\ee
with $d\omega_i$ being the canonically normalized $\left(\frac{1}{4\pi i}\oint_{A_i}d\omega_i=
\delta_{i,j}\right)$ holomorphic
1-differentials on the reduced
Riemann surface ${\why}(\lambda)$.
Here $ H_{i}(\lambda)$ are
polynomials of degrees $n-2$ (the degree is exact because each such
polynomial has exactly one zero on each cycle $A_j$ for $j\ne i$).

Rewriting now (\ref{Dvdual1}) as $\frac{\d{\cal F}_0}{\d S_i}=\oint_{B_i}dS$,
together with (\ref{canoDV}), we reconstruct the SW system~\cite{WS1}
whose prepotential is the planar limit ${\cal F}_0$ of the 1MM free energy.
From this formula and (\ref{canoDV}) we obtain the celebrated expression relating
${\cal F}_0$ and the period matrix $\tau_{i,j}$ of the Riemann surface:
\be
\frac{\d^2{\cal F}_0}{\d S_i\d S_j}=\oint_{B_i}d\omega_j\equiv \tau_{i,j}.
\label{tau}
\ee

\paragraph{Matrix integral as a Whitham--Krichever system.}

We now show that ${\cal F}_0$ enjoys additional geometrical structures arising when
differentiating w.r.t. times $t_k$ and $t_0$ thus accommodating the whole SWWK system.

We consider variations of the potential, i.e., variations
w.r.t. (Whitham) times $t_k$. We then obtain instead of (\ref{v2})
\be\label{v3}
\delta dS=-{1\over 2}{\delta \left((V'_m)^2(\lambda)-\frac{P_{m+t-1}(\lambda)}
{\prod_{\beta=1}^t(\lambda-a_\beta)}\right)
\over M_{m-\frac{s-t}2}(\lambda){\tilde y}(\lambda)}d\lambda
\ee
while (\ref{v1}) still holds. Repeating the arguments of the previous
paragraph, we conclude that the zeroes of $M_{m-\frac{s-t}2}(\lambda)$ cancel from the
denominator and, therefore, the variation may have pole only at
$\lambda=\infty$, or $\eta=0$, i.e., at the puncture. In order
to find this pole, we
use (\ref{v1}), which implies that
$dS=M_{m-\frac{s-t}2}(\lambda){\tilde y}(\lambda)d\lambda\to (V'_m(\lambda)
+O({1\over\lambda}))d\lambda$ and, therefore, the variation of $dS$ at large
$\lambda$ is completely determined by the variation of $V'_m(\lambda)$.
Parameterizing $V(\lambda)=\sum^{m+1}_{k=1}t_k{\lambda^{k}}$, we obtain
that
\be
\label{ds}
{\partial dS\over\partial S_i}=d\omega_i,\qquad
{\partial dS\over\partial t_k}=2d\Omega_k
\ee
with the behavior at infinity
\be
\label{*Omega*}
d\Omega_k=k\left(\xi^{-k-1}+O(1)\right)d\xi,\ \hbox{for} \ \xi\to 0,\ k>0.
\ee
This holds up to a linear combination of holomorphic differentials.
The normalization of $d\Omega_k$ is fixed by the condition
\be
\label{dVSi}
\frac{\partial S_i}{\partial t_k}=
\frac{1}{2\pi i}
\oint_{A_i}d\Omega_k=0\ \forall \ i,k,\ \ \
\hbox{or }\ \  {\d S_i\over\d V(\lambda)}=0\ \ i=1,\ldots,n-1.
\ee

The derivatives of $dS$ w.r.t. the times are
\be
2d\Omega_k \equiv {\d dS \over \d t_k}
={ H_{n+k-1}(\lambda) d\lambda\over \why(\lambda)},
\label{omes}
\ee
and normalization conditions (\ref{dVSi})
together with the asymptotic expansion
\bea
2d\Omega_k(\lambda)|_{\lambda\to\infty}&=&{k\lambda^{k-1}d\lambda}
+O(\lambda^{-2})d\lambda=
\nonumber
\\
&=&{k\lambda^{k-1}d\lambda}+\sum_{m=1}^{\infty}
c_{km}\lambda^{-1-m}d\lambda.
\label{ass}
\eea
fixes uniquely the coefficients of the
corresponding polynomials $ H_{n+k-1}$
of degrees $n+k-1$.

Equations (\ref{ds}) together with normalization conditions (\ref{dVSi})
determine the SWWK system~\cite{Kri1} (see also \cite{RG,MarMir98}) whose
prepotential is ${\cal F}_0$.
For this, we apply the formula similar to
(\ref{*b2*}) allowing now variations of the potential~$V(\lambda)$.
We then obtain (see (\ref{*b2*})--(\ref{Dvdual}))
\bea
\frac{\partial {\cal F}_0}{\partial t_k}&=&
-\frac1{4\pi i}\oint_{\cal D}d\lambda
\frac{\partial y(\lambda)}{\partial t_k}\cdot h(\lambda)
-\frac1{4\pi i}\oint_{\cal D}d\lambda y(\lambda) {\lambda^k}
=-\sum_{i=1}^{n-1} \frac{\partial S_i}{\partial t_k}(\zeta_i-\zeta_n)+
\frac12\res_{\lambda=\infty}(\lambda^{k}dS)
\nonumber
\\
&=&\frac12\res_{\lambda=\infty}(\lambda^{k}dS)\equiv \frac12 v_k,\quad k=1,\dots,m,
\label{v41}
\eea
which is again the standard formula of the SWWK theory.

We now consider the variation of $dS$ w.r.t. the hard wall parameters $a_\beta$.
Note first that
because $S_i$ and $t_0$ are now independent variables, together with $t_k$ and $a_\beta$,
we must demand
\be
\label{daSi}
\frac{\partial S_i}{\partial a_\beta}=
\frac{1}{2\pi i}
\oint_{A_i}\frac{\d y(\lambda)}{\d a_\beta}d\lambda=0\ \forall \ i,\beta,\quad\hbox{and}
\quad \frac{\d V'(\lambda)}{\d a_\beta}=0.
\ee

Performing the variation of (\ref{1mamocu}) w.r.t. $a_\beta$, we obtain
\be
y(\lambda)\delta_{a_\beta}y(\lambda)=2\frac{\delta_{a_\beta}P_{m+t-1}(\lambda)}
{\sqrt{\prod_{\gamma=1}^t(\lambda-a_\gamma)}}+\frac{P_{m+t-1}(\lambda)}
{(\lambda-a_\beta)\sqrt{\prod_{\gamma=1}^t(\lambda-a_\gamma)}}=\frac{{\overline P}_{m+t-1}(\lambda)}
{(\lambda-a_\beta)\sqrt{\prod_{\gamma=1}^t(\lambda-a_\gamma)}},
\label{v4}
\ee
so, again, using considerations on cancelation of zeros of the polynomial $M(\lambda)$, we have
\be
\frac{\d y(\lambda)}{\d a_\beta}=\frac{{\tilde P}_{n-1}(\lambda)}{(\lambda-a_\beta)\why(\lambda)}
\equiv d\Omega_{a_\beta}(\lambda),
\label{v5}
\ee
where $d\Omega_{a_\beta}(\lambda)$ is a (noncanonical) meromorphic 1-form with the only
second-order pole at $\lambda=a_\beta$ and with all its $A$-cycle integrals vanishing due to
condition (\ref{daSi}). To find the quadratic residue of this form at $\lambda=a_\beta$,
we need the most singular contribution when performing the
variation of $y(\lambda)=M(\lambda)\ty(\lambda)$ w.r.t. $a_\beta$; this contribution
comes from the partial derivative in $a_\beta$, and we have
\be
d\Omega_{a_\beta}(\lambda)= \frac12 M(a_\beta)\frac{\sqrt{\prod_{\alpha=1}^s(a_\beta-\mu_\alpha)}}
{\sqrt{\prod_{\gamma=1 \atop \gamma\ne\beta}^t(a_\beta-a_\gamma)}}\frac{d\lambda}{(\lambda-a_\beta)^{3/2}}
+O\left(\frac{1}{\sqrt{\lambda-a_\beta}}\right)d\lambda;\qquad \oint_{A_i}d\Omega_{a_\beta}(\lambda)=0 \ \forall i.
\label{asdO}
\ee

\paragraph{On $t_0$-dependence of the prepotential.}

The last property of ${\cal F}_0$ to be verified is the behavior w.r.t. $t_0$.
From the SWWK theory, we expect that
\be
\label{dut0}
{\d {\cal F}_0\over\d t_0}=\int^{\infty_+}_{\infty_-}dS
\ee
and $\frac{\d dS}{\d t_0}=d\Omega_0$ where $d\Omega_0$ is the Abelian differential of the
third kind with two simple poles at two infinities:
\be\label{bip}
\res_{\infty}d\Omega_0=-\res_{\infty_-}d\Omega_0=-1.
\ee
The integral in (\ref{dut0}) is however divergent and needs regularization.

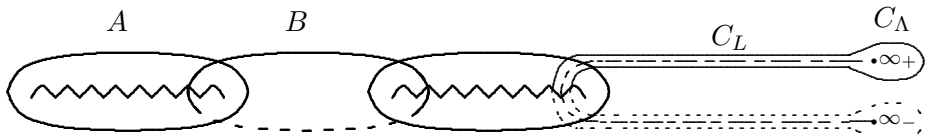
\begin{figure}[tb]
%\hspace*{2cm}
%\epsfysize=6cm
\vskip .2in
\setlength{\unitlength}{0.8mm}%
\begin{picture}(0,40)(15,15)
\thicklines
\curve(60,30, 62,33.5, 65,35, 70,36, 80,36.5, 90,36, 95,35, 98,33.5, 100,30, 98,26.5, 95,25, 90,24, 80,23.5, 70,24, 65,25, 62,26.5, 60,30)
\curve(64,29, 66,31, 68,29, 70,31, 72,29, 74,31, 76,29, 78,31, 80,29, 82,31, 84,29, 86,31, 88,29, 90,31, 92,29, 94,31, 96,29)
\put(80,40){\makebox(0,0)[rb]{$A$}}
\curve(120,30, 122,33.5, 125,35, 130,36, 140,36.5, 150,36, 155,35, 158,33.5, 160,30, 158,26.5, 155,25, 150,24, 140,23.5, 130,24, 125,25, 122,26.5, 120,30)
\curve(124,29, 126,31, 128,29, 130,31, 132,29, 134,31, 136,29, 138,31, 140,29, 142,31, 144,29, 146,31, 148,29, 150,31, 152,29, 154,31, 156,29)
\put(110,40){\makebox(0,0)[rb]{$B$}}
\curve(92,26.5, 90,30, 92,33.5, 95,35, 100,36, 110,36.5, 120,36, 125,35, 128,33.5, 130,30, 128,26.5)
{\curvedashes[1mm]{0,1,2}
\curve(92,33.5, 90,30, 92,26.5, 95,25, 100,24, 110,23.5, 120,24, 125,25, 128,26.5, 130,30, 128,33.5)}
\put(204,35){\circle*{1}}
\put(204,25){\circle*{1}}
{\thinlines
\curvedashes[1mm]{1,3,2}
\curve(204,35, 157,35, 154,34, 153,33, 152,30, 153,27, 154,26, 157,25, 204,25)
\curvedashes[1mm]{5,1}
\curve(204,35, 157,35, 154,34, 153,33, 152,30, 153,27, 154,26, 157,25, 204,25)}
\thinlines
\put(180,37){\makebox(0,0)[cb]{$C_L$}}
\put(207,40){\makebox(0,0)[cb]{$C_\Lambda$}}
\put(208,35){\makebox(0,0)[cc]{${\scriptstyle \infty_+}$}}
\put(208,25){\makebox(0,0)[cc]{${\scriptstyle \infty_-}$}}
\curve(153.5,30, 154,31.8, 154.5,32.8, 155.5,33.5, 157.5,34, 164,34, 199,34, 200,34, 201,33.5, 205,32, 209,32, 210,32.3, 211,32.8, 212,34, 212.3,35)
\curve(150.5,30, 151,32.2, 151.5,33.7, 152.5,35, 154.5,36, 162,36, 199,36, 200,36, 201,36.5, 205,38, 209,38, 210,37.7, 211,37.2, 212,36, 212.3,35)
\curvedashes[0.5mm]{2,1}
\curve(153.5,30, 154,28.2, 154.5,27.2, 155.5,26.5, 157.5,26, 164,26, 199,26, 200,26, 201,26.5, 205,28, 209,28, 210,27.7, 211,27.2, 212,26, 212.3,25)
\curve(150.5,30, 151,27.8, 151.5,26.3, 152.5,25, 154.5,24, 162,24, 199,24, 200,24, 201,23.5, 205,22, 209,22, 210,22.3, 211,22.8, 212,24, 212.3,25)
\end{picture}
%\centerline{\epsfxsize=11.5cm\epsffile{wdvv11.eps}}
\caption{Cuts in the $\lambda$-, or ``eigenvalue," plane for the
planar limit of 1MM. The eigenvalues are supposed to be
located ``on" the cuts. We add the logarithmic
cut between two copies of the infinity on two sheets of the
hyperelliptic Riemann surface in order to calculate the derivative
w.r.t. the variable $t_0$ and let $C_L$ denote the corresponding integration contour.}
\label{fi:cuts}
\end{figure}

We know the derivative $\d{\cal F}_0/\d t_0$,
(\ref{Pi0}), which is equal to the integral
$\oint_{\cal D}\log|\lambda-q|dS-V(q)$ with the
reference point $q$ to be chosen on the last, $n$th, cut, while the
expression itself does not depend on the actual local position of the
reference point. We choose it to be
$\mu_{2n}$---the
rightmost point of the cut. We can then invert the contour
integration over the support $\cal D$ to the integral along the contour
that runs first along the upper side of the logarithmic cut
from $\mu_{2n}$ to a regularization point $\Lambda$, then
over the circle $C_\Lambda$ of large radius $|\Lambda|$ and then back over
the lower side of the logarithmic cut in the complex plane. In order to
close the contour on the hyperelliptic Riemann surface under consideration,
we must add the integration over the corresponding contour on the {\it second}
sheet of the surface
as shown in Fig.~\ref{fi:cuts}; we
let $C_L$ denote the completed integration contour, and
it is easy to see that such an
additional integration just double the value of the integral.

All the singularities appearing at the upper
integration limit (i.e., at the point~$\Lambda$) are exactly cancelled
by the contribution coming when
integrating the expression $dS\log(\lambda-\mu_{2n})$
along the circle $C_\Lambda$; in fact, the latter can be
easily done, the result is $-2\pi i\bigl(S(\Lambda)-(S(\mu_{2n}))_+\bigr)$, where the
function $S(\lambda)$ is the (formal)
primitive of $dS$ (which includes the logarithmic term),
and we project it to the strictly polynomial part.
Using the large-$\lambda$ asymptotic expansion
of the differential $dS$,
\be
\label{asympt}
dS(\lambda)|_{\lambda\to\infty}=V'(\lambda)d\lambda
+{t_0\over\lambda}d\lambda+O(\lambda^{-2})d\lambda,
\ee
we obtain that $(S(\mu_{2n}))_+$ just cancels the term $V(\mu_{2n})$,
and we eventually find that
\bea
{\d F\over\d t_0}&=&\frac12\left(\oint_{C_L}\log(\lambda-\mu_{2n})dS-2V(\mu_{2n})\right)=
\nonumber
\\
&=&2\pi i\left(\int_{\mu_{2n}}^\Lambda dS-S(\Lambda)\right),
\label{dfdt0}
\eea
where $C_L$ is the contour in Fig.~\ref{fi:cuts},
which by convention encircles the logarithmic cut between two infinities
on two sheets of the Riemann surface and passes through the last, $n$th, cut.
We have therefore proved that the planar limit ${\cal F}_0$ of the 1MM free energy
is the SWWK prepotential, or semiclassical tau-function.

We now introduce the (complete) set of canonical variables $\{S_i,\
i=1,\dots, n-1;\ t_0;\ t_k,\ k=1,\dots,m\}$, which we uniformly denote $\{t_I\}$
(in what follows, Latin capitals indicate any quantity from this set).
From (\ref{canoDV}), (\ref{bip}), and (\ref{omes}), we then obtain the
general relation
\be
\label{dOI}
{\d dS\over\d t_I}\equiv d\Omega_I=
{ H_{I}(\lambda)d\lambda\over\why(\lambda)},
\ee
where $ H_{I}(\lambda)$ are polynomials.

Asymptotic formulas (\ref{bip}) and (\ref{ass}) imply
that {\it derivatives\/} of all the quantities $d\Omega_I$
w.r.t. any parameter
are regular at infinity and may have singularities
{\it only\/} at the branching points $\mu_i$ of reduced
Riemann surface (\ref{tyred}).

Note again that these derivatives are purely geometrical and does not
depend, in our setting, on the nature of the branching points: whether they
come as hard walls or as ``dynamical'' branching points.

\section{Derivatives of the free energy and geometry \label{s:geometry}}

\subsection{Second derivatives of the free energy \label{ss:second}}

\paragraph{Bergmann bidifferential and 2-point resolvent. \label{ss:BPQ}}
Calculating the one-point resolvent in the preceding section
required the knowledge of the first derivatives of the
matrix model free energy. We now turn to the two-point
resolvents involving second derivatives.
Here, instead of differentials with some prescribed
properties of holomorphicity, the main object we need is a
{\it bi}differential on a Riemann surface~$\Sigma_g$: the
{\it Bergmann kernel} (canonically normalized bidifferential in Fay's terminology)
which is the double derivative of the logarithm of the
Prime form $E(P,Q)$
such that it is symmetrical in its arguments $P,Q\in \Sigma_g$
and has the only singularity at the coinciding arguments with the
behavior (see~\cite{Fay})
\be
B(P,Q)=\left(\frac{1}{(\xi(P)-\xi(Q))^2}+O(1)\right)d\xi(P)d\xi(Q),
\label{*Bergmann*}
\ee
in some local coordinate $\xi(P)$. As it stands, we can add to (\ref{*Bergmann*}) any bilinear
combination of Abelian 1-differentials $d\omega_i$; we fix the normalization claiming
vanishing all the integrals over $A$-cycles of $B(P,Q)$:
\be
\oint_{A_i}B(P,Q)=0,\ \hbox{for}\ i=1,\dots,g,
\label{*vanish*}
\ee
and, due to the symmetricity property, the integral may be taken over any of the
variables $P$ or $Q$.

We now show that the Bergmann kernel generates the
differentials $d\Omega_k$. Bipole differential
(\ref{bip}) can be expressed through the Prime form as
\be
\label{bp}
d\Omega_{0} = d\log{E(P,\infty_+)\over E(P,\infty_-)}
\ee
The primitive of differential (\ref{bp}) (which we need in what follows)
then obviously develops the logarithmic cut
between the points of two infinities on the Riemann surface.

Using (\ref{4.5}), (\ref{v41}) and (\ref{ds}), we obtain
\bea
W_0(\lambda,\mu)d\lambda\,d\mu&=&
-\sum^{\infty}_{j= 0}{d\mu d\lambda\over
\lambda^{j+1}} \frac{\d}{\d t_j}W_0(\mu) =
-\frac12 \frac{\d}{\d V(\lambda)}V'(\mu)d\lambda\,d\mu+
\frac12 \sum^{\infty}_{j=0}
\frac{d\lambda\,d\mu}{\lambda^{j+1}}\frac{\d}{\d t_j}y(\lambda)
\nonumber
\\
&=&-\frac12 \frac{d\lambda\,d\mu}{(\lambda-\mu)^2}+
\frac12 \sum^{\infty}_{j=0}\frac{d\lambda}{\lambda^{j+1}}d\Omega_j(\mu).
\label{BPQ}
\eea
Because $d\Omega_j(\mu)=\left(\pm j\mu^{j-1}+O(1/\mu^2)\right)d\mu$ for
$\mu\to\infty_{\pm}$, \ $j>0$, we conclude that the infinite sum
$\frac12 \sum^{\infty}_{j=0}\frac{d\lambda}{\lambda^{j+1}}d\Omega_j(\mu)$
develops the second-order pole $\frac{d\lambda\,d\mu}{2(\lambda-\mu)^2}$
in the case where $\lambda$ and $\mu$ are on the same (physical) sheet; this
pole is then canceled by the first term, and the second-order pole
$-\frac{d\lambda\,d\mu}{2(\lambda-\mu)^2}$ for $\lambda$ and $\mu$ being on
different sheets. In the latter case, this pole is doubled by the first term,
and together with the evident normalizing conditions
that follow immediately from (\ref{BPQ}),
\be\label{a-per}
\oint_{A_i}W_0(\lambda,\mu)d\mu=
\oint_{A_i}W_0(\lambda,\mu)d\lambda=0,
\ee
we finally come to the formula for the 2-point resolvent,
\be
W_0(\lambda ,\mu )d\lambda d\mu =\frac{\partial W_0(\lambda )}
{\partial V(\mu )}d\lambda d\mu =-B(P,Q^*),
\label{*two-loop*}
\ee
and, correspondingly,
\be
\frac{\d y(\lambda)}{\d V(\mu)}=-\frac12 (B(P,Q)-B(P,Q^*)),
\label{dVy}
\ee
where we have introduced the $*$-involution between two sheets
of our hyperelliptic curve, so that $Q^*$ denotes
the image of $Q$ under this involution.
The only singularity of (\ref{*two-loop*}), for a fixed point $P$ on a physical
sheet, is at the point $Q\to P^*$
on the unphysical sheet with $\mu(Q)=\lambda(P^*)=\lambda(P)$.

In what follows, we try as much as possible to express everything in terms of invariant
quantities (the Bergmann bidifferentials and their primitives) on the reduced
Riemann surface (\ref{tyred}) and in terms of the WK differential $dS$.
As we shall see, the $1/N$-expansion can be constructed basically in terms of just these
characteristics. However, we somewhere need the explicit expressions for
$B(P,Q)$.

We need a technically simpler expression for the
one-differential $d{\cal E}(\lambda ,\mu )$, which is
the primitive of the Bergmann kernel
$B(\lambda ,\mu )$ w.r.t. the argument $\mu $.\footnote{To be more precise, it is
primitive of the antisymmetrized Bergmann kernel
$\frac12\bigl(B(\lambda,\mu)-B(\lambda^*,\mu)\bigr)$.}
Obviously, it is a single-valued differential of $\lambda $ with
zero $A$-periods on the reduced Riemann surface and is multiple-valued function of $\mu $,
which undergoes jumps equal to $d\omega_i(\lambda )$ when the variable $\mu $ passes
along the cycle $B_i$:
\bea
&&d{\cal E}\left(P,Q+\oint_{B_j}\right)=2\pi i dw_j(P)+d{\cal E}(P,Q),
\nonumber
\\
&&d{\cal E}\left(P,Q+\oint_{A_j}\right)=d{\cal E}(P,Q).
\nonumber
\eea
In Sec.~\ref{s:genus}, we use the explicit representation
for the kernel $d{\cal E}(\lambda,\mu)$ in the hyperelliptic case:
\be
\label{*dE}
d{\cal E}(\lambda ,\mu )=\frac{\why(\mu )d\lambda }{(\lambda -\mu )\why(\lambda )}-
\sum_{i=1}^{n-1}\frac{ H_{i}(\lambda )d\lambda }{\why(\lambda )}
\oint_{A_i}d\xi \frac{\why(\mu )}{(\xi-\mu )\why(\xi)}.
\ee
We take $d{\cal E}(\lambda ,\mu )$ to be antisymmetric w.r.t. the involution
$\lambda\to\lambda^*$ between the physical and unphysical sheets, so it has the simple
pole with the residue one at $\lambda=\mu$ and the simple pole with the residue minus one
at $\lambda^*=\mu$ having vanishing $A$-periods in the variable $\lambda$. Then
$d{\cal E}(\lambda ,\mu )=\int_{\mu_0}^{\mu}B(\lambda,\xi)-\int_{\mu_0}^{\mu^*}B(\lambda,\xi)$,
with $\mu_0$ the reference point on which there is no dependence.

Note that formula (\ref{asdO}) expressing the derivative of $y(\lambda)$ w.r.t.
$a_\beta$ is proportional to $d{\cal E}(\lambda,a_\beta)$ and, moreover, can be
conveniently written as the residue at the branching point $a_\beta$:
\be
\frac{\d y(\lambda)d\lambda}{\d a_\beta}\equiv
d\Omega_{a_\beta}(\lambda)=\res_{\mu=a_\beta}\left(B(\lambda,\mu)y(\mu)\right),
\label{resab}
\ee
whereas at a branching point $\mu_i$ ($i=1,\dots,2n$), we have
\be
2B(\lambda,[\mu_i])=\lim_{\mu\to\mu_i}\frac{d{\cal E}(\lambda,\mu)}{\sqrt{\mu-\mu_i}}.
\label{B->E}
\ee

In terms of $B(P,[\mu_i])$, we have one of the Rauch variational identities \cite{Rauch}:
\be
\frac{\d}{\d \mu_i}B(P,Q)=\frac12 B(P,[\mu_i])B(Q,[\mu_i]), \quad i=1,\dots,2n.
\label{Rauch}
\ee

\paragraph{Mixed second derivatives. Normalizing conditions for $W_h(\lambda)$.}

Another set of relations follows from the general properties of the Bergmann kernel, and
can be also derived directly from the formulas of Sec.~\ref{s:1MM}. To this end,
we apply the mixed derivatives
$\partial /\partial V(\mu )$ and $\d/\d S_i$ to the planar limit of the free energy ${\cal F}_0$. On
one hand, $\d{\cal F}_0/\d S_i=\oint_{B_i}dS$ and using that
$dS(\lambda)=y(\lambda)d\lambda=(V'(\lambda)-2W_0(\lambda))d\lambda$,
${\partial V'(\lambda)\over\partial V(\mu)}=-{1\over (\lambda-\mu)^2}$ and
formula (\ref{*two-loop*}), we have
\be
\oint_{B_i}\frac{\partial (dS(\lambda ))}{\partial V(\mu )}d\mu
=\oint_{B_i}\left(2B(P,\mu)-\frac1{(\lambda -\mu )^2}dpd\mu\right)
=\oint_{B_i}2B(P,\mu).
\label{abc}
\ee
On the other hand, acting by derivatives in
the opposite order, we first obtain
$\partial {\cal F}_0/\partial V(\mu )d\mu=W_0(\mu )d\mu=(V'(\mu )-y(\mu ))d\mu$
and then $\d (V'(\mu )-y(\mu ))d\mu/\d S_i =2d\omega_i(\mu)$,
or, in the coordinate-free notation,
one of the classic identities (see, e.g.,~\cite{Fay}):
\be
{1\over 2\pi i}\oint_{B_i}B(P,Q)=d\omega_i(Q)
\label{*B-Bcycle*}
\ee
This means that
\be
\label{dsdt}
{\partial dS(\mu) \over \d S_i}=\left[\oint_{B_i}{\d dS(\lambda )\over\d V(\mu)}d\lambda\right]d\mu=
\oint_{B_i}{\d dS(\mu)\over\d V(\lambda )}d\lambda
\ee
where the {\it both} integrals are taken over the variable $\lambda $.
Now, as $dS=y(\lambda)d\lambda$ is the generating function for the variables
$\xi_a\equiv
\{\{b_i\},\{\mu_\alpha\},M_\alpha^{(i)}\}$, given by (\ref{ty}), (\ref{M}),
and (\ref{M2}) (recall that $M_\alpha^{(i)}$ are just $i$th order derivatives of $M(\lambda )$
at $\lambda =\mu_\alpha$) giving their dependence
on $S_i$ and $t_k$, one concludes that similar relation for the first
derivatives holds for each of these variables.
Indeed, multiplying (\ref{dsdt}) by ${1\over
y(\mu)}$ one then can bring $\mu$ successively to $\mu_\alpha$'s, $\lambda_i$'s
and $\infty$ to pick up pole terms with different $\xi_a$ and prove that
\be
{\d \xi_a\over\d S_i}=\oint_{B_i} {\d \xi_a\over\d V(\lambda )}d\lambda
\ee
which becomes true also for {\em any} function depending on a {\em finite} number of
``local'' variables $\xi_a$ in the case of arbitrary potential. For instance, it does not
hold for $S_j$ for which $\d S_j/\d S_i=\delta_{i,j}$ whereas $\d S_j/\d V(\lambda)\equiv 0$.
However, one can {\em never} express $S_i$ as a function
of a {\em finite} number of ``local'' variables $\xi_a$.

Since, as we show in sec.~\ref{s:genera}, the free energy ${\cal F}_h$ at any order
of $1/N$-expansion depends only on a finite number of local variables, which are
the branching points $\mu_\alpha$ and a finite number of moments $M_\alpha^{(i)}$,
and $W_h(\lambda)$ is then expressed exclusively in
terms of a finite number of derivatives $\frac{\d \mu_\alpha}{\d V(\lambda)}$ and
$\frac{\d M_\alpha^{(k)}}{\d V(\lambda)}$, we have
\be\label{G}
{\d {\cal F}_h\over\d S_i}=\oint_{B_i} {\d {\cal F}_h\over\d V(\lambda )}d\lambda
=\oint_{B_i} W_h(\lambda)d\lambda
\quad\forall h\hbox{ and } i=1,\dots,g,
\ee
together with the condition
\be
\label{A-cycle-F}
\oint_{A_i}\frac{\d \F_h}{\d V(\lambda)}d\lambda\equiv
\oint_{A_i}W_h(\lambda)d\lambda=0\quad \forall h\ge1\hbox{\ and\ for\ }i=1,\dots,g,
\ee
that follows directly from normalization condition (\ref{*vanish*}).

In the same way, we can obtain the defining relation for the $t_0$-derivative
\be
\frac{\d {\cal F}_h}{\d t_0}=\int_{\infty_-}^{\infty_+}\frac{\d {\cal F}_h}{\d V(\xi)}d\xi,
\label{Z}
\ee
which again holds for any $h$ and requires regularization only for $h=0$ since
higher corrections in $h$ are regular at infinities.

\subsection{Residue formula and WDVV equations  \label{ss:WDVV}}

The Witten--Dijkgraaf--Verlinde--Verlinde (WDVV)
equations~\cite{Wit90,LGMT}
is the systems of algebraic equations~\cite{wdvvg}
\be
\label{WDVV}
\F_I{\F}_J^{-1}\F_K = \F_K{\F}_J^{-1}\F_I, \quad \forall\ I,J,K
\ee
on the third derivatives
\be
\label{matrF}
\|{\F}_{I}\|_{JK}=
{\d^3\F\over\d t_I\,\d t_J\,\d t_K} \equiv\F_{IJK}
\ee
of some function $\F (\{t_I\})$. These equations can be often
interpreted as associativity relations in some algebra for
an underlying topological theory.

That the 1MM free energy satisfy the WDVV equations
was proved in \cite{ChMMV}, where it was shown that
the multicut solution of the 1MM satisfies the WDVV
equations as a function of {\em canonical} variables identified with
the periods and residues of the generating meromorphic one-form $dS$
\cite{Kri1}. The method to prove it consists of two steps.
The first, most difficult, step is to find the residue formula for third
derivatives (\ref{matrF}) of the 1MM free energy. Then, using an
associativity, one immediately proves that the
free energy of multi-support solution satisfies
the WDVV equations if we keep the number of independent variables
to be {\it equal\/} to the number of ``dynamical''
branching points. That is, the presence of the hard-wall branching points
results in the reducing the total dimension of the WDVV system.

\paragraph{Residue formula}

Because all the quantities $d\Omega_I$ (\ref{dOI}) depend {\it entirely\/}
on reduced hyperelliptic Riemann surface (\ref{tyred}), their
derivatives w.r.t. {\it any parameter\/} must be expressed through
the derivatives w.r.t. the positions of the branching points $\mu_i$.
We skip most of details referring the reader to \cite{Kri1,ChMMV}.

The formula for the third derivative
$\d^3\F_0/(\d t_I\d t_J\d \mu_\alpha)\equiv \F^{(0)}_{IJ\alpha}$
easily follows from (\ref{Rauch}) when $t_I$ and $t_J$ are times of the potential, and one
should use the Riemann bilinear identities in the case of other canonical variables.

We eventually have
\be
\label{Fay}
\F^{(0)}_{IJ\alpha}=\res_{\mu_\alpha}\bigl(d\Omega_I\d_\alpha\Omega_J\bigr)
={ H_{I}(\mu_\alpha)
 H_{J}(\mu_\alpha)\over \prod_{j\ne\alpha}
(\mu_\alpha-\mu_j)}.
\ee

Completing the calculation
of the third derivative needs just inverting the dependence on the branching
points therefore finding $\d \mu_\alpha/\d t_K$. (Note that, obviously,
$\d a_\beta/\d t_K\equiv 0$.)
Differentiating
expressions (\ref{1mamocu}), (\ref{ty}) w.r.t. $t_K$, we obtain
for (\ref{ds})
\bea
&&{\d dS(\lambda)\over \d t_K}=
{ H_{K}(\lambda)d\lambda\over \why(\lambda)}=
\nonumber
\\
&&\quad\quad=\frac12 M(\lambda)\sum_{\alpha=1}^{s}
{{\tilde y}(\lambda)\over
(\lambda-\mu_\alpha)}{\d \mu_\alpha\over \d t_K}d\lambda+
{\d M(\lambda)\over\d t_K}\tilde y(\lambda)d\lambda.
\label{diff1}
\eea
The derivative of the polynomial $M(\lambda)$ is obviously polynomial
and regular at $\lambda=\mu_\alpha$. Multiplying (\ref{diff1})
by $\sqrt{\lambda-\mu_\alpha}$ and setting $\lambda=\mu_\alpha$,
we immediately obtain
\be
\label{diff2}
\frac{\d \mu_\alpha}{\d t_K}=\frac{ H_{K}(\mu_\alpha)}
{M(\mu_\alpha)\prod_{\gamma=1\atop\gamma\ne\alpha}^s
(\mu_\alpha-\mu_\gamma)},
\ee
with {\em only} ``dynamical'' branching points in the product in the denominator.
Combining this with (\ref{Fay}), we come to the desired residue formula for the
third derivative w.r.t. the canonical variables $t_I$:
\bea
{\d^3 \F_0\over \d t_I\d t_J\d t_K} &=&
\sum_{\alpha=1}^{s}{ H_I(\mu_{\alpha})
 H_J(\mu_{\alpha}) H_K(\mu_{\alpha})\over
M(\mu_\alpha)
\prod_{\gamma=1\atop\gamma\ne\alpha}^s(\mu_{\alpha}-\mu_{\gamma})^2
\prod_{\beta=1}^t(\mu_{\alpha}-a_\beta)}=
\nonumber
\\
&=&\sum_{i=1}^{s}\res_{\mu_{\alpha}}
{d\Omega_Id\Omega_Jd\Omega_K\over d\lambda dy}
=\sum_{\alpha=1}^{2n}\res_{\mu_i}
{d\Omega_Id\Omega_Jd\Omega_K\over d\lambda
dy}=
\res_{d\lambda=0}{d\Omega_Id\Omega_Jd\Omega_K\over d\lambda dy}.
\label{resgen}
\eea
We are able to pass to summation over {\em all} branching points in the last line
because the residues at $\lambda=a_\beta$ just vanishes and in the very last
expression we assume zeros of the differential $d\lambda$ on the reduced Riemann
surface (\ref{tyred}).

\paragraph{Proof of WDVV equations}

Given residue formula (\ref{resgen}), the proof of WDVV equations
(\ref{WDVV}) can be done by checking associativity of the algebra
of differentials $d\Omega_I$ with multiplication modulo
$\sqrt{\prod_{\alpha=1}^s(\lambda-\mu_\alpha)}d\lambda$.
This algebra is reduced to the algebra of polynomials
$ H_I(\lambda)$ with multiplication modulo $\prod_{\alpha=1}^s(\lambda-\mu_\alpha)$,
which is correctly defined and
associative. The basis of the algebra of $ H_I(\lambda)$
obviously has dimension $s$ and is given, e.g., by monomials of
the degrees $0,1,\dots,s-2,s-1$.

Another proof pertains to solving the system of linear equations \cite{BMRWZ,MaWDVV}.
Defining
\be
\label{phi}
\phi_I^{\alpha}\equiv
{ H_I(\mu_{\alpha})
\over
M^{1/3}(\mu_\alpha)
\prod_{\gamma=1\atop\gamma\ne\alpha}^s(\mu_{\alpha}-\mu_{\gamma})^{2/3}
\prod_{\beta=1}^t(\mu_{\alpha}-a_{\beta})^{1/3}},
\ee
we reduce (\ref{resgen}) to the form
\be\label{Fphi3}
\F^{(0)}_{IJK}=\sum_{\alpha=1}^s \phi_I^{\alpha}\phi_J^{\alpha}\phi_K^{\alpha}.
\ee

We also demand the determinant of $\phi_{I}^{\alpha}$ to be nonzero:
\be
\det\limits_{I\alpha}\| \phi_{I}^{\alpha}\|\ne0.
\label{det}
\ee
This nondegeneracy
holds automatically  in the case where we have no hard walls and
choose the canonical variables to be $S_i$, $t_0$, and first $n$ times $t_k$
($k=1,\dots,n$) of the potential~\cite{ChMMV2}.\footnote{From the definition (\ref{phi}),
this condition stems, up to obviously nonvanishing
Vandermonde determinant factors, to the nondegeneracy of the matrix $\sigma$ (\ref{Q}).
Indeed, if $\det \sigma$ vanishes, then there exists a polynomial $P(\lambda)$ of degree less or
equal $n-2$ such that
$$
\int_{\mu_{2i-1}}^{\mu_{2i}}
\frac{P(\lambda)}{\why(\lambda)}d\lambda=0\quad\hbox{for}
\quad i=1,\dots,n-1.
$$
This necessarily implies that $P(\lambda)$ has at least one zero
at each of the intervals $(\mu_{2i-1},\mu_{2i})$; otherwise
the combination under the integral sign is sign definite and the
integral cannot vanish. The polynomial $P(\lambda)$ must then
have at least $n-1$ zero and, having the degree not exceeding $n-2$,
must therefore vanish.} \
In the case with hard walls, however, no such consideration works, and
this determinant {\em may} vanish on subdomains of parameters of the problem,
so we must consider condition (\ref{det}) as an {\em additional} restriction,
besides the matching condition  claiming that {\em the number of canonical variables
must coincide with the number of ``dynamical'' branching points of the solution}.
Recall that we need at least three canonical variables in order to have
a nontrivial WDVV system.

We now let $\F^{(0)}_I$ denote
the matrix with the entry $JK$ equal to $\F^{(0)}_{IJK}$. The WDVV equations
(\ref{WDVV}) follow now immediately from (\ref{Fphi3}) and nondegeneracy condition
(\ref{det}).

\section{Higher genus contributions \label{s:genera}}

The solution $W_1(\lambda )$ to the loop equations in the multicut case
was first found by
Akemann~\cite{Ak96} and the universality of critical behavior of the
corresponding correlation functions was demonstrated in~\cite{AkAm}.
Akemann also managed to integrate the 1-point resolvent
to obtain the free energy $\F_1$ in the two-cut case.
The genus-one partition function in the generic multi-cut
case was proposed in \cite{Kos,DST}, where it was observed that the Akemann formula
coincides with the correlator of twist fields, computed by Al.Zamolodchikov
\cite{Zam}. The function $F_1$ in the general multicut case was found by
integrating $W_1(\lambda )$ in \cite{Chekh}, \cite{ChMMV2}.

\subsection{The iterative solution of the loop equation}\label{iterate}

\paragraph{Inverting the integral operator.}

We now determine
higher genus contributions. We do this iteratively by inverting
the genus expanded loop equation (\ref{4.11}).
Our strategy will be to
construct an integral operator $\widehat {d{\cal E}}$
inverse to the integral operator
$\widehat{K}-2W_0(\lambda)$.

Acting with this operator on the both sides of loop equation (\ref{4.11}) we
recursively obtain $W_h(\lambda )$ for all genera like all the
multi-point resolvents of the same genus can be obtained from $W_h(\lambda )$ merely
by applying the loop insertion operator $\dV$. Another ingredient of the construction
will be the action of the loop insertion operator within our technique.

However, there is a subtlety: the operator $\widehat{K}-2W_0(\lambda)$ has
zero modes and is not invertible. Therefore, solution to the loop equation
is determined up to an arbitrary combination of these zero modes.
Moreover, in the case of hard walls, the kernel of $\widehat {d{\cal E}}$
does not coincide with the one of $\widehat{K}-2W_0(\lambda)$ and must be
discussed separately.

From representation (\ref{plan}) and (\ref{*loop3*}), we have that the action of
$\widehat{K}-2W_0$ on $W_h(\lambda)$ is
\bea
\bigl(\widehat {K}-2W_0(\lambda)\bigr)W_h(\lambda)&=&
\bigr[V'(\lambda)W_h(\lambda)\bigl]_- -\bigl(V'(\lambda)-y(\lambda)\bigr)W_h(\lambda)
\nonumber
\\
&=&\bigr[V'(\lambda)W_h(\lambda)\bigl]_- -\left[\bigl(V'(\lambda)-y(\lambda)\bigr)W_h(\lambda)\right]_-
\nonumber
\\
&=&\bigr[y(\lambda)W_h(\lambda)\bigl]_-.
\label{K}
\eea
Because both $W_h(\lambda)$ and $y(\lambda)$ are odd functions under the involution
$\lambda\to\lambda^*$ interchanging the physical and unphysical sheets, we conclude that
the combination $y(\lambda)W_h(\lambda)$, being an even function under this
involution, is a meromorphic function on the physical sheet, that is,
it is a rational function having poles only at the branching points and at infinity.

It is now easy to see that the integral operator
\be
\label{1*}
\widehat{d{\cal E}}\left(f\right)\equiv
\oint_{{\cal C}_{\cal D}}
\frac{d\mu }{2\pi i}\,\frac{d{\cal E}(\lambda ,\mu )}
{d\lambda }\frac{1}{y(\mu )}
\cdot f(\mu )
\ee
with the kernel $d{\cal E}(\lambda ,\mu )$ from (\ref{*dE})
is an inverse for the operator $\widehat{K}-2W_0(\lambda)$.
Indeed, representing $f(\mu)$ in form (\ref{K}), we have
\bea
&{}&\oint_{\lambda>{\cal C}_{\cal D}}
\frac{d\mu }{2\pi i}\,\frac{d{\cal E}(\lambda ,\mu )}
{d\lambda }\frac{1}{y(\mu )}\bigl[y(\mu)W_h(\mu)\bigr]_-
=\oint_{\lambda>{\cal C}_{\cal D}}
\frac{d\mu }{2\pi i}\,\frac{d{\cal E}(\lambda ,\mu )}
{d\lambda }\frac{1}{y(\mu )}\bigl(y(\mu)W_h(\mu)-P_{m-2}(\mu)\bigr)
\nonumber
\\
&{}&\quad
=\oint_{\lambda>{\cal C}_{\cal D}}
\frac{d\mu }{2\pi i}\,\frac{d{\cal E}(\lambda ,\mu )}
{d\lambda }\left(W_h(\mu)-\frac{P_{m-2}(\mu)}{y(\mu)}\right)
\nonumber
\\
&{}&\quad
=W_h(\lambda)
+\oint_{{\cal C}_{\infty_+}}
\frac{d\mu }{2\pi i}\,\frac{d{\cal E}(\lambda ,\mu )}
{d\lambda }W_h(\mu)
-\oint_{\lambda>{\cal C}_{{\cal D}}}
\frac{d\mu }{2\pi i}\,\frac{d{\cal E}(\lambda ,\mu )}{d\lambda }
\frac{P_{m-2}(\mu)}{y(\mu)},
\label{inv1}
\eea
where $P_{m-2}(\mu)$ is a polynomial.
Here, we indicated by $\lambda>{\cal C}_{\cal D}$ that the point $\lambda$ lies
outside the integration contour in the physical sheet and, when pulling the contour of
integration to $\infty_+$, only the residue at $\mu=\lambda$ contributes
in the first integral and, by the normalization conditions,
this residue is exactly $W_h(\lambda)$. The remaining integral at infinity vanishes
because the integrand is regular as $\mu\to\infty$.
Next, by virtue of explicit formula (\ref{*dE}),
we see that because the last term is a rational function on the physical sheet that is regular at
the branching points, integrating it over ${\cal C}_{\cal D}$ obviously gives zero, so the
whole action of $\widehat{d{\cal E}}$ on $(\widehat {K}-2W_0)W_h(\lambda)$
just reconstructs $W_h(\lambda)$.

Note also that under the action on $[y(\mu)W_h(\mu)]_-$, due to the involution properties,
we can replace the integration over ${\cal C}_{\cal D}$ just by evaluating residues at
$\mu=\mu_i$ in formula (\ref{1*}) having
\be
\label{1**}
\widehat{d{\cal E}}\left(f\right)(\lambda)d\lambda=
\oint_{{\cal C}_{\{\mu_i\}}}
\frac{d\mu }{2\pi i}\,d{\cal E}(\lambda ,\mu )
\frac{1}{y(\mu )}
\cdot f(\mu ),
\ee
where we let ${\cal C}_{\{\mu_i\}}$ denote the union of contours encircling
all the branching points.

Because the operator $\widehat{d{\cal E}}$ obeys the property
\be\label{normG}
\oint_{A_i}\widehat{d{\cal E}}(f)(\lambda )d\lambda \equiv 0
\ee
it also enjoys integrability conditions (\ref{A-cycle-F}). Therefore, one has to
solve the loop equations inverting $\widehat{K}-2W_0$ exactly with
$\widehat d{\cal E}$.

\paragraph{Zero modes. The modified loop equation.}

Note that now, in contrast to the case without hard walls, the zero-mode content of the operators
$\widehat{K}-2W_0$ and $\widehat{d{\cal E}}$ is different.

It is clear from representation (\ref{K}) that the
zero modes of the operator $\widehat{K}-2W_0$ when acting on the functions $W_h(\lambda)$ with
asymptotic behavior (\ref{Winf}) are spanned by $P_{\frac{s-t}2-2}(\lambda)/\ty(\lambda)$ and constitute the
linear space of dimension $\frac{s-t}2-1$. In addition, {\em all the simple-pole functions} proportional
to $\frac1{\lambda-a_\beta}$ are annihilated by the operator $\widehat {d{\cal E}}$. Altogether, it
gives (in the case $s>t$) the total space of zero modes of dimension $\frac{s-t}2-1+t=n-1=g$, and the
total number of zero modes under the action of the combination of
the operators $\widehat{d{\cal E}}\circ (\widehat{K}-2W_0)$ is therefore
equal to the number $g$ of Abelian differentials,
whereas the freedom to add these differentials is completely fixed by normalization condition (\ref{A-cycle-F}).

A crucial observation is that when acting by the operator $\widehat{d{\cal E}}$ on the both sides of
loop equation (\ref{4.11}), {\em the terms with the partial derivatives of $F_h$ w.r.t. $a_\beta$
vanish}, the action of $\widehat{K}-2W_0$ on $W_h(\lambda)$ is inverted, and the remained expression
just becomes
\be
W_h(\lambda)=\widehat{d{\cal E}}\left(\sum_{h'=1}^{h-1}
W_{h'}(\lambda)W_{h-h'}(\lambda)+\frac{\partial }{\partial V(\lambda)}W_{h-1}(\lambda)\right),
\label{loopmod}
\ee
with $W_h(\lambda)$ now automatically satisfying normalization conditions (\ref{A-cycle-F}).

It follows from the properties of the operator $\widehat{d{\cal E}}$ that,
for all $s>0$ at $h>0$ and for $s\ge3$ at $h=0$, the $s$-point resolvent
$W_h(\lambda_1,\dots,\lambda_s)$ possesses the gradation w.r.t. involution
between variables of the physical and unphysical sheets:
\be
W_h(\lambda_1,\dots,\lambda_j,\dots\lambda_s)=
-W_h(\lambda_1,\dots,{\lambda^*_j},\dots\lambda_s).
\label{gradation}
\ee
In particular, $W_h(\lambda,\lambda)$ as well as any product
$W_{h}(\lambda)W_{h'}(\lambda)$ in the r.h.s. of (\ref{loopmod}) are then
rational functions of $\lambda$ for any $h,h'>0$. Moreover,
$W_0(\lambda,\lambda)=B(\lambda,\lambda^*)$ is again a rational function of
$\lambda$ having poles of the
second order at the point $\lambda=\mu_i$ of merging of the two sheets.

However, in order to have loop equation (\ref{4.11}), not (\ref{loopmod}), we must demand
additional conditions to be satisfied by the matrix-model free energy. In contrast to
conditions (\ref{G}) satisfied by any function of local variables, these conditions
impose restrictions; indeed, it is required for the parts of loop equation (\ref{4.11}) proportional
to $\frac1{\lambda-a_\beta}$ to match. That is, the constructed loop means $W_h(\lambda)$
must satisfy
\bea
\frac{\d{\cal F}_h}{\d a_\beta}&=&
\oint_{{\cal C}_{a_\beta}}\frac{d\xi}{2\pi i}
\left(-\bigl[y(\xi)W_h(\xi)\bigr]_-+\sum_{h'=1}^{h-1}
W_{h'}(\xi)W_{h-h'}(\xi)+\frac{\partial }{\partial V(\xi)}W_{h-1}(\xi)\right)
\nonumber
\\
&=&
\oint_{{\cal C}_{a_\beta}}\frac{d\xi}{2\pi i}
\left(-y(\xi)W_h(\xi)+\sum_{h'=1}^{h-1}
W_{h'}(\xi)W_{h-h'}(\xi)+\frac{\partial }{\partial V(\xi)}W_{h-1}(\xi)\right),
\label{match-a}
\eea
because the residue at $\xi=a_\beta$ of the positive part $\bigl[y(\xi)W_h(\xi)\bigr]_+$
is obviously zero.

In two next paragraphs, we solve recurrently modified loop equation (\ref{loopmod}). We then integrate it
to produce the free energy term ${\cal F}_h$ in the next section and, after obtaining the answer,
must verify conditions (\ref{match-a}). Note that the two normalization conditions,
(\ref{G}) and (\ref{match-a}), leave no room for adding
terms independent on $V(\xi)$ and dependent either on $S_i$ or on $a_\beta$ alone. And, as we show in
the last section, in the special case of ${\cal F}_1$, condition (\ref{match-a}) fixes
the function of $a_\beta$ to be added to the free energy.

\paragraph{Diagrammatic technique.}

The above considerations provide a basis for the {\em diagrammatic representation} for resolvents, which
almost literally repeats the diagrammatic technique for the 1MM without hard walls~\cite{Ey}.
Let us represent the one-form $d{\cal E}(p,q)$ as the vector directed
from $p$ to $q$, the three-point vertex as the dot in which
we assume the integration over $q$, \ $\bullet\equiv\oint\frac{dq}{2\pi i}\frac{1}{2y(q)}$, and the
Bergmann bidifferential $B(p,q)$ as a nonarrowed edge connecting points $p$ and $q$. The graphic representation
for a solution of (\ref{loopmod}) then looks as follows.
Representing the multiresolvent $W_{h'}(p_1,\dots,p_s)$ as the block with $s$ external legs and with the index $h'$,
we obtain~\cite{Ey}
\be
\hbox{
\begin{picture}(190,50)(10,20)
\thicklines
\put(40,40){\oval(20,20)}
\thinlines
\put(20,40){\line(1,0){10}}
\put(20,40){\circle*{2}}
\put(20,44){\makebox(0,0)[cb]{$\lambda $}}
\put(40,40){\makebox(0,0)[cc]{$h$}}
\put(55,40){\makebox(0,0)[lc]{$=\sum\limits_{h'=1}^{h-1}$}}
\put(80,40){\vector(1,0){13}}
\put(80,40){\circle*{2}}
\put(95,40){\circle*{4}}
\put(80,44){\makebox(0,0)[cb]{$\lambda $}}
\put(93,44){\makebox(0,0)[cb]{$\mu $}}
\put(95,40){\line(1,1){7}}
\put(95,40){\line(1,-1){7}}
\thicklines
\put(109,54){\oval(20,20)}
\put(109,26){\oval(20,20)}
\put(109,54){\makebox(0,0)[cc]{$h{-}h'$}}
\put(109,26){\makebox(0,0)[cc]{$h'$}}
\thinlines
\put(120,40){\makebox(0,0)[lc]{$+$}}
\put(130,40){\vector(1,0){13}}
\put(130,40){\circle*{2}}
\put(145,40){\circle*{4}}
\put(130,44){\makebox(0,0)[cb]{$\lambda $}}
\put(143,44){\makebox(0,0)[cb]{$\mu $}}
\qbezier(145,40)(150,49)(157,49)
\qbezier(145,40)(150,31)(157,31)
\thicklines
\put(166,40){\oval(25,25)}
\put(166,40){\makebox(0,0)[cc]{$h{-}1$}}
\put(185,40){\makebox(0,0)[cc]{$,$}}
\end{picture}
}
\label{Bertrand}
\ee
which provides the basis for the diagrammatic representation for $W_h(p_1,\dots,p_s)$.
It can be formulated as a set of the following diagrammatic rules~\cite{Ey}.

The multiresolvent $W_h(p_1,\dots,p_s)$ is presented as a finite sum of {\em all possible} connected
graphs with $h$ loops and $s$ external legs such that
\begin{itemize}
\item  only three-valent internal vertices are allowed
(the total number of edges is then $2s+3h-3$, and we assume $s\ge1$ for $h\ge1$ and $s\ge3$ for $h=0$);
all the vertices are indexed by the corresponding coordinate variables $q$ on the Riemann surface $\why(\lambda)$.
\item We have propagators (edges) which either connect internal vertices (can start and terminate
at the same vertex) or come as an external leg to an internal vertex.
\item These propagators are of two sorts: there are {\em arrowed} propagators and {\em nonarrowed} propagators.
As above, we set the 1-forms $d{\cal E}(p,q)$ into the correspondence to the arrowed propagators and the variables
$p$ and $q$ correspond to the respective vertices at which this edge starts and terminates. We have the
Bergmann bidifferential $B(p,q)$ corresponding to each nonarrowed edge that starts and terminates at the
vertices labeled by the respective variables $p$ and $q$ and if it is the same vertex, then $B(p,p^*)$
corresponds to this propagator.
\item The arrangement of propagators is subject to restrictions.
We have exactly $2k+s-2$ arrowed edges; these edges must constitute the {\em maximum rooted tree
subgraph} with {\em all arrows directed from the root}. Different choices of the tree subgraph for the
same graph are considered different cases of the construction, and we must consider all possible cases.
The root of the tree subgraph is always
a selected (one and for all) {\em external} leg, say, $p_1$
(the choice is arbitrary due to the symmetry of $W_k(p_1,\dots,p_s)$) and the maximality of the
tree subgraph means that it contains no loops (including tadpoles;
the arrowed propagators cannot therefore start and terminate at the same vertex), is connected, and
for every internal vertex there is exactly one arrowed propagator that terminates at this vertex. We call
this propagator the {\em incoming} propagator for this vertex. There can be either one, or two, or no {\em outgoing} arrowed
propagators at an internal vertex of the graph.
\item Each tree subgraph establishes the relation of {\em partial ordering} on the set of internal vertices
of the graph; among two vertices one {\em precedes} to the other if there exist a naturally oriented
path in the rooted arrowed tree subgraph that starts at the first vertex and terminates at the second vertex.
At each internal vertex we place the integration $\oint_{{\cal C}^{(q)}_{{\cal D}}}\frac{dq}{2\pi i}\frac{1}{2y(q)}$
over the variable indexing this vertex; the integration goes
along the contour that is about the domain ${\cal D}$ and the {\em arrangement} of the integration contours
at different vertices is prescribed by the arrowed subtree: the closer is a vertex to the root, the more outer
is the integration contour.
\item All internal
nonarrowed propagators are allowed to connect {\em only} vertices that are in the partial ordering
relation to one another; it can be the same vertex. All external propagators except the one that is the
root of the tree subgraph are nonarrowed, and we have the Bergmann bidifferentials $B(q,p_k)$ \ $(k=2,\dots,s)$
corresponding to these edges.
\item All the integration contours over the $A$-cycles in formula (\ref{*dE}) for $d{\cal E}(\lambda,\mu)$
are assumed to be outside all the integration contours over the variables indexing the internal vertices whereas
all the external points $p_i$, $i=1,\dots,s$ are assumed to be outside all the integration contours
${\cal C}^{(q)}_{{\cal D}}$ (as well as outside all the contours of integration over the $A$-cycles).
\end{itemize}

We now demonstrate the consistency of this diagrammatic technique by calculating the action of loop insertion
operator $\d/\d V(r)$ on its elements following \cite{CEy}.

We first calculate the action of $\d/\d V(r)$ on $B(P,Q)$ thus producing
an analogue of formula (\ref{resgen}).
Using (\ref{Rauch}), we represent this action through
the action of partial derivatives in $\mu_\alpha$ \ ($\alpha=1,\dots,s$; only the dynamical
branching points participate on this stage)
subsequently calculating the latter from relation (\ref{dVy}).
Let
$$
y(x)dx|_{x\to\mu_\alpha}=y([\mu_\alpha])\sqrt{x-\mu_\alpha}dx+O(\sqrt{x-\mu_\alpha})^3dx.
$$
Then, since
$$
\left.\frac{\d y(p)dp}{\d V(r)}\right|_{p\to\mu_\alpha}\simeq
-\frac12 y([\mu_\alpha])\frac{dp}{\sqrt{p-\mu_\alpha}}\frac{\d\mu_\alpha}{\d V(r)},
$$
we have
\be
\frac{\d \mu_\alpha}{\d V(r)}=\frac{2B([\mu_\alpha],r)}{y([\mu_\alpha]),}
\label{dmul}
\ee
and, therefore
\be
\frac{\d}{\d V(r)}B(P,Q)=\sum_{\alpha=1}^{s} \frac{B(P,[\mu_\alpha])B(Q,[\mu_\alpha])B(r,[\mu_\alpha])}{y([\mu_\alpha])}.
\label{Y.4}
\ee
This is apparently just formula
(\ref{resgen}) upon setting all variables $I,J,K$ to be the times of the potential with subsequent
summation with $p^{-i-1}$, $q^{-j-1}$, and $r^{-k-1}$ for $p$, $q$, and $r$ being the
variables on the physical sheet.

However, now we represent expression (\ref{Y.4}) in the different way. Instead of differentiating $y(\xi)$
in the denominator as we did in formula (\ref{resgen}), we integrate one of the Bergmann kernels, which
gives the 1-differential $d{\cal E}(Q,\xi)$. Note that the local
variable in the vicinity of $\mu_\alpha$ is $\xi(x)=\sqrt{x-\mu_\alpha}$, this gives the additional factor $1/2$, so
we have
$$
\frac{\d}{\d V(r)}B(P,Q)=\sum_{\alpha=1}^{s} \res_{\mu_\alpha}\frac{B(P,\xi(x))d{\cal E}(Q,\xi(x))B(r,\xi(x))}{2y(\xi(x))d\xi(x)},
$$
and we can again {\em add} residues at the hard wall points $a_\beta$ because they are just vanish here,
so our final expression is
\be
\frac{\d}{\d V(r)}B(P,Q)=\sum_{i=1}^{2n} \res_{\mu_i}\frac{B(P,\xi(x))d{\cal E}(Q,\xi(x))B(r,\xi(x))}{2y(\xi(x))d\xi(x)}.
\label{Y.5}
\ee
The price for preserving $y(\xi(x))d\xi(x)=dS$ in the denominator is that we have lost the explicit permutational
symmetry for the three terms in the numerator, which was apparent in (\ref{resgen}). We however pay this price
for the possibility to develop the comprehensive diagrammatic technique.
From this relation, it obviously follows that
\be
\frac{\d}{\d V(r)}d{\cal E}(Q,P)=\sum_{i=1}^{2n} \res_{\mu_i}
\frac{d{\cal E}(\xi(x),P)d{\cal E}(Q,\xi(x))B(r,\xi(x))}{2y(\xi(x))d\xi(x)},
\label{Y.6}
\ee
and the last quantity to evaluate is
\be
\frac{\d}{\d V(r)}\frac{1}{2y(p)}=-\frac{B(p,r)}{2y^2(p)}.
\label{Y.7}
\ee

Note that the point $P$ in (\ref{Y.6}) is outside the integration contour.
Multiplying the both sides of (\ref{Y.6}) by $1/(2y(P))$,
using (\ref{Y.7}), and pushing the integration contour through the point $P$, we observe that the contribution of the
simple pole of $d{\cal E}(\xi(x),P)$ at the point $\xi(x)=P$ {\em cancels exactly} the variation of $1/(2y(P))$!
We therefore attain the prescribed contour ordering and can graphically present the action of $\d/\d V(r)$ as
\be
\begin{picture}(240,20)(10,40)
\thicklines
\put(5,40){\makebox(0,0)[cc]{$\frac{\d}{\d V(r)}$}}
\put(20,40){\makebox(0,0)[cc]{$Q$}}
\put(25,40){\vector(1,0){15}}
\put(45,40){\makebox(0,0)[cc]{$P$}}
\put(55,40){\makebox(0,0)[cc]{$=$}}
\put(65,40){\makebox(0,0)[cc]{$Q$}}
\put(70,40){\vector(1,0){10}}
\put(80,40){\circle*{2}}
\put(80,40){\vector(1,0){10}}
\put(95,40){\makebox(0,0)[cc]{$P,$}}
\put(80,40){\line(0,1){10}}
\put(80,55){\makebox(0,0)[cc]{$r$}}
\put(125,40){\makebox(0,0)[cc]{$\frac{\d}{\d V(r)}$}}
\put(140,40){\makebox(0,0)[cc]{$Q$}}
\put(145,40){\line(1,0){15}}
\put(165,40){\makebox(0,0)[cc]{$P$}}
\put(175,40){\makebox(0,0)[cc]{$=$}}
\put(185,40){\makebox(0,0)[cc]{$Q$}}
\put(190,40){\vector(1,0){10}}
\put(200,40){\circle*{2}}
\put(200,40){\line(1,0){10}}
\put(215,40){\makebox(0,0)[cc]{$P$}}
\put(200,40){\line(0,1){10}}
\put(200,55){\makebox(0,0)[cc]{$r$}}
\put(225,40){\makebox(0,0)[cc]{$\equiv$}}
\put(235,40){\makebox(0,0)[cc]{$Q$}}
\put(240,40){\line(1,0){10}}
\put(250,40){\circle*{2}}
\put(260,40){\vector(-1,0){10}}
\put(265,40){\makebox(0,0)[cc]{$P\ .$}}
\put(250,40){\line(0,1){10}}
\put(250,55){\makebox(0,0)[cc]{$r$}}
\end{picture}
\label{variation}
\ee
In the second case, we have the freedom to choose
on which of edges to set the arrow. Recall, however, that as we have had a nonarrowed propagator
stretched between the points $P$ and $Q$, these points were to be
ordered before, as prescribed by the diagrammatic technique.
That is, if ``$P$ preceded $Q$'' we must choose the first variant and if ``$Q$ preceded $P$''
we must choose the second variant of arrows arrangement.

\subsection{Inverting the loop insertion operator. Free energy  \label{s:free}}

\paragraph{The $H$-operator. \label{ss:H}}

We now introduce the operator that is inverse to loop insertion operator (\ref{4.6}).
Let\footnote{This definition works well when acting on 1-forms regular at infinities. Otherwise (say, in the case of
$W_0(p)$), the integral in the third term must be regularized, e.g., by replacing it by the contour integral around the logarithmic
cut stretched between two infinities as was done in Sec.~\ref{ss:prepotential}.}
\be
H\cdot=\frac12\res_{\infty_+}V(x)\cdot\ -\frac12\res_{\infty_-}V(x)\cdot\
-t_0\int_{\infty_-}^{\infty_+}\cdot\ -\sum_{i=1}^{n-1}S_i\oint_{B_i}\cdot\ .
\label{H}
\ee
The arrangement of the integration contours is as in Fig.~\ref{fi:cuts}. We calculate the action of $H$ on the Bergmann
bidifferential $B(x,q)$ using again the Riemann bilinear identities.
We first note that as $B(x,q)=\d_xd{\cal E}(q,x)$, we can evaluate residues at infinities
by parts. Then, since $d{\cal E}(q,x)$ is regular at infinities, we substitute $2y(x)+2t_0/x$ for $V'(x)$ as $x\to\infty_+$
and $-2y(x)+2t_0/x$ for $V'(x)$ as $x\to\infty_-$ thus obtaining
\bea
&&-\res_{\infty_+}\left(y(x)+\frac{t_0}{x}\right)d{\cal E}(q,x)dx+\res_{\infty_-}\left(-y(x)+\frac{t_0}{x}\right)d{\cal E}(q,x)dx
\nonumber
\\
&&\qquad \Bigl.-t_0d{\cal E}(q,x)\Bigr|_{x=\infty_-}^{x=\infty_+}-\sum_{i=1}^{n-1}S_i\oint_{B_i}B(q,x).
\label{Z.1}
\eea
Whereas the cancelation of terms containing $t_0$ is obvious, it remains only to take the combination of residues
at infinities involving $y(x)$. For this, we cut the surface along $A$- and $B$-cycles taking into account the residue at
$x=q$. The boundary integrals on two sides of the cut at $B_i$ then differ by $d{\cal E}(q,x)-d{\cal E}(q,x+\oint_{A_i})=0$,
while the integrals on the two sides of the cut at $A_i$ differ by  $d{\cal E}(q,x)-d{\cal E}(q,x+\oint_{B_i})=\oint_{B_i}B(q,x)$,
and the boundary term therefore becomes
$$
\sum_{i=1}^{n-1}\oint_{A_i}y(x)dx\oint_{B_i}B(q,\xi),
$$
which exactly cancels the last term in (\ref{Z.1}). Only the contribution from the pole at $x=q$ then survives, and this
contribution is just $-y(q)$. We have therefore proved that
\be
H\cdot B(\cdot,q)=-y(q)dq.
\label{HB}
\ee

\paragraph{Free energy.}

We now act by $H$ on $W_k(\cdot)$ subsequently evaluating the action of loop insertion operator (\ref{4.6}) on the
result. Note first that the only result of action of $\d/\d V(p)$ on the operator $H$ itself are derivatives
$\d V(x)/\d V(p)=-1/(p-x)$ (and recall that by definition $|p|>|x|$, i.e., instead of evaluating residues at infinities one should
take residues at $x=p$), which gives
\be
\frac{\d}{\d V(p)}\left(H\cdot W_h(\cdot)\right)=W_h(p)+H\cdot W_h(\cdot,p).
\label{Z.2}
\ee
For the second term, due to the symmetry of $W_h(p,q)$, we may choose now
the point $p$ as the root of all the tree subgraphs. Then,
the operator $H$ always acts on $B(\cdot,\xi)$ where $\xi$ are integration variables of internal vertices. However, if this vertex
is an innermost (i.e., it has no outgoing arrowed edges), then the 1-form $y(\xi)d\xi$ arising under the action of $H$
(\ref{HB}) cancels the corresponding 1-form in the denominator of the integrand, which becomes therefore regular
at the branching point giving zero contribution.
If this vertex has an outgoing arrowed edge, say $d{\cal E}(\xi,\rho)$
(it can be at most one outgoing arrowed edge as the third edge must be
external), then, again, we can push the integration contour for $\xi$ through the one for $\rho$;
the only contribution comes only from the pole at $\xi=\rho$. The value of the residue is however
{\em doubled}, and we come to the
following graphical representation for the action of the operator $H$:
\be
\begin{picture}(240,30)(10,35)
\thicklines
\put(25,40){\makebox(0,0)[cc]{$Q$}}
\put(30,40){\vector(1,0){10}}
\put(40,40){\circle*{2}}
\put(40,40){\vector(1,0){10}}
\put(55,40){\makebox(0,0)[cc]{$P$}}
\put(40,40){\line(0,1){10}}
\put(40,55){\makebox(0,0)[cc]{$H\cdot$}}
\put(65,40){\makebox(0,0)[cc]{$=$}}
\put(75,40){\makebox(0,0)[cc]{$-$}}
\put(85,40){\makebox(0,0)[cc]{$Q$}}
\put(90,40){\vector(1,0){15}}
\put(110,40){\makebox(0,0)[cc]{$P$}}
\put(116,40){\makebox(0,0)[cc]{;}}
\put(135,40){\makebox(0,0)[cc]{$Q$}}
\put(140,40){\vector(1,0){10}}
\put(150,40){\circle*{2}}
\put(150,40){\line(1,0){10}}
\put(165,40){\makebox(0,0)[cc]{$P$}}
\put(150,40){\line(0,1){10}}
\put(150,55){\makebox(0,0)[cc]{$H\cdot$}}
\put(175,40){\makebox(0,0)[cc]{$=$}}
\put(185,40){\makebox(0,0)[cc]{$0.$}}
\end{picture}
\label{chopping}
\ee
For $H_q\cdot W_k(q,p)=H_q\cdot \frac{\d}{\d V(q)}W_k(p)$, we obtain that for each {\em arrowed} edge on which the action of
$\d/\d V(r)$ produces the new vertex, the inverse action of $H_q\cdot$ just give the factor $-1$ and for each {\em nonarrowed} edge
on which the action of $\d/\d V(r)$ produces the new vertex, the inverse action of $H_q\cdot$ just gives zero. As the total number
of arrowed edges is exactly $2h-1$ for every graph contributing to the sum of diagrams with one external edge and containing $h$ loops,
we obtain that
$$
H_q\cdot W_h(q,p)=-(2h-1)W_h(p).
$$
Therefore, combining with (\ref{Z.2}), we just obtain
\be
\frac{\d}{\d V(p)}\left(H_q\cdot W_h(q)\right)=-(2-2h)\frac{\d}{\d V(p)}{\cal F}_h,
\label{Z.3}
\ee
and, since all the dependence on filling fractions and $t_0$ is fixed by the claim that the answer depends only on $\mu_i$ and
derivatives of $M(p)$ of finite orders at the branching points, we conclude that
\be
{\cal F}_h=\frac{1}{2h-2}H\cdot W_h.
\label{fin}
\ee
We have therefore expressed the free energy through the one-loop resolvent.
Using this final answer, we can calculate {\em all} ${\cal F}_h$ except the contribution at $h=1$ (the torus approximation).
In the case without hard walls, the latter
was calculated by a direct integration in~\cite{Chekh} (see the refined version in \cite{ChMMV2}), and we generalize this
calculation to the hard wall case in Sec.~\ref{s:genus}.
All other orders can be calculated consistently. For this,
we need to introduce one new vertex $\circ$\hskip-0.85ex$\cdot$
in which we place the new integration term:
\be
\hbox{$\circ$\hskip-0.85ex$\cdot$}\sim
\oint_{{\cal C}^{(\xi)}_{\{\mu_i\}}}\frac{d\xi}{2\pi i}\frac{\int_{\mu_i}^\xi y(s)ds}{y(\xi)}.
\label{key}
\ee
This is because, although the integral term $\int_{Q_0}^\xi y(s)ds$ is nonlocal, its constant part
$\int_{Q_0}^{\mu_i}$, having the improper involution symmetry, drops out of the residue in the
1MM case,
and we can integrate only in the neighborhood of each branching point $\mu_i$ separately. That is, we need only the local expansion
of this integral as $\xi\to\mu_i$. Then, say, the genus two contribution is provided by the sum of three diagrams
\be
\hbox{
\begin{picture}(240,20)(10,30)
\thicklines
\put(15,40){\makebox(0,0)[cc]{$2\cdot {\cal F}_{2}$}}
\put(30,40){\makebox(0,0)[cc]{$=$}}
\put(38,40){\makebox(0,0)[cc]{$2$}}
\put(45,40){\circle{3}}
\put(45,40){\circle*{1.5}}
%\put(45,40){\makebox(0,0)[cc]{$\cdot$}}
\curve(45.4,41.2, 46.4,43.4, 48.4,44.6, 50,45, 51.6,44.6, 53.6,43.4, 54.6,41.2)
\curve(45.4,38.8, 46.4,36.6, 48.4,35.4, 50,35, 51.6,35.4, 53.6,36.6, 54.6,38.8)
%\put(50,40){\circle{10}}
\put(55,40){\circle*{2}}
\put(53,45){\vector(1,0){0}}
\put(56.5,40){\vector(1,0){7}}
\put(65,40){\circle*{2}}
\curve(65.4,41.2, 66.4,43.4, 68.4,44.6, 70,45, 71.6,44.6, 73.6,43.4, 74.6,41.2, 74.8,40,
 74.6,38.8, 73.6,36.6, 71.6,35.4, 70,35, 68.4,35.4, 66.4,36.6, 65.4,38.8)
%\put(70,40){\circle{10}}
%%
\put(85,40){\makebox(0,0)[cc]{$+2$}}
\put(95,40){\circle{3}}
\put(95,40){\circle*{1.5}}
\put(106.5,40){\oval(20,20)[r]}
%\put(105,50){\circle*{2}}
\put(105,48.5){\vector(0,-1){17}}
\put(105,50){\circle*{2}}
\put(105,30){\circle*{2}}
\curve(95.2,41.4, 96.8,45, 100,48.2, 103.4,49.8)
\curve(95.2,38.6, 96.8,35, 100,31.8, 103.4,30.2)
\put(100,48.5){\vector(1,1){0}}
%\put(105,30){\circle*{2}}
%%
\put(125,40){\makebox(0,0)[cc]{$+$}}
\put(140,40){\circle*{2}}
\curve(139.6,41.2, 138.6,43.4, 136.6,44.6, 135,45, 133.4,44.6, 131.4,43.4, 130.4,41.2, 130.2,40,
 130.4,38.8, 131.4,36.6, 133.4,35.4, 135,35, 136.6,35.4, 138.6,36.6, 139.6,38.8)
\put(148.5,40){\vector(-1,0){7}}
\put(150,40){\circle{3}}
\put(150,40){\circle*{1.5}}
\put(151.5,40){\vector(1,0){7}}
\put(160,40){\circle*{2}}
\curve(160.4,41.2, 161.4,43.4,163.4,44.6, 165,45, 166.6,44.6, 168.6,43.4, 169.6,41.2, 169.8,40,
 169.6,38.8, 168.6,36.6, 166.6,35.4, 165,35, 163.4,35.4, 161.4,36.6, 160.4,38.8)
\put(173,39){\makebox(0,0)[cc]{.}}
\end{picture}
}
\label{F2}
\ee

Note also that ${\cal F}_0$ written in form (\ref{FS}) obviously can be presented as
\be
{\cal F}_0=-\frac{1}{2}H_{\rm reg}\cdot W_0
\ee
due to (\ref{Dvdual}) and (\ref{Pi0}) and by virtue of relations (\ref{G}) and (\ref{Z}).
Recall that in the case of ${\cal F}_0$, the integral between $\infty_-$ and $\infty_+$
must be regularized as in Sec.~\ref{ss:prepotential}.

Postponing the consideration of the case ${\cal F}_1$ to the next section, let us show here that
$H\cdot W_1$ is constant. Indeed,
$$
H\cdot W_1=\sum_{i=1}^{2n}\res_{\mu_i}\frac{\int_{\mu_i}^\xi y(s)ds}{y(\xi)}B(\xi,\xi^*)d\xi,
$$
and the first term has simple zero with
residue $2/3$ or $2$ (depending on which branching point---hard wall or dynamical---we have at $\mu_i$)
to be compensated by the double pole of $B(\xi,\xi^*)\simeq \frac{1}{4(\xi-\mu_\alpha)^2}$,
and the total answer is then just the constant equal to $(2s/3+2t)/4$.

\paragraph{Scaling relation. \label{ss:scaling}}

We now explain the origin of relation (\ref{Z.3}). Indeed, recall that for any functional ${\cal F}$ of a finite
number of ``local'' variables, which are in our case the branching points $\mu_i$ and
the moments $M_i^{(k)}$, we have relations (\ref{G}) and (\ref{Z}) that state that
$\oint_{B_i}\frac{\d {\cal F}_h}{\d V(\xi)}d\xi=\d{\cal F}_h/\d S_i$ and
that $\int_{\infty_-}^{\infty_+}\frac{\d {\cal F}_h}{\d V(\xi)}d\xi=\d{\cal F}_h/\d t_0$.
Relation (\ref{Z.3}) is then equivalent to the formula
\be
\left[\sum_{k=1}^\infty t_k\frac{\d}{\d t_k}+t_0\frac{\d}{\d t_0}+\sum_{i=1}^{n-1}S_i\frac{\d}{\d S_i}
+\hbar\frac{\d}{\d \hbar}\right]{\cal F}=0,
\label{base}
\ee
which is just a (``kinematic'') property of {\em every} integral of form (\ref{Xap2.1}). Indeed, let us fix the total number
of eigenvalues to be $N$ and the partial numbers of eigenvalues on intervals of distribution to be $N_i$. Then, $t_0=\hbar N$,
$S_i=\hbar N_i$, and we have the exponential term $\frac1\hbar \sum_k t_k \tr X^k$. If we just keep $N$, $N_i$, and all the
combinations $t_k/\hbar$ fixed and scale {\em only} the formal expansion parameter $\hbar$, we immediately come
to (\ref{base}).

Note that in the case of 1MM without hard walls, the
operator in (\ref{base}) has also the sense of
the scaling transformations under which all vertices are multiplied
by the same scaling factor $\rho$ and all the propagators are multiplied by $\rho^{-1}$. The action of derivatives in $t_0$ and $S_i$
results in multiplying all index loops (faces of the fat graph) by $\rho$. Therefore, for any graph, the total factor is
$$
\rho^{\hbox{\scriptsize \#\ vertices -- \#\ edges + \#\ faces}}=\rho^{2-2h},
$$
and it is exactly canceled by the scaling of the formal expansion parameter $\hbar\to\rho\hbar$.

The above relation, being purely kinematic, pertains to any matrix integral irrespectively to the domain of
eigenvalues, etc.; to the best of our knowledge, this is the first time when it found a technical application.

\paragraph{On checking the consistency conditions for $h\ge2$.}

The consistency condition for $h=0$ are satisfied by construction. We verify these conditions
in the case $h=1$ in the next section and consider here only the case $h\ge2$ in which we have now the explicit
representation for the free energy ${\cal F}_h$.

A simple, but rather naive, consideration pertains to that the whole expression for ${\cal F}_h$ with
$h\ge2$ can be treated as a functional of $y(\mu)$. Then,
\bea
\oint_{{\cal C}_{a_\beta}}\frac{d\lambda}{2\pi i}
[y(\lambda)W_h(\lambda)]_-&=&
\oint_{{\cal C}_{a_\beta}}\frac{d\lambda}{2\pi i}
y(\lambda)\frac{\d {\cal F}_h}{\d V(\lambda)}=
\oint_{{\cal C}_{a_\beta}}\frac{d\lambda}{2\pi i}
\oint_{\lambda>{\cal C}_{\cal D}}\frac{d\mu}{2\pi i}
y(\lambda)\frac{\d y(\mu)}{\d V(\lambda)}\frac{\delta {\cal F}_h}{\delta y(\mu)}
\nonumber
\\
&{}&\quad\quad=\oint_{{\cal C}_{a_\beta}}\frac{d\lambda}{2\pi i}
\oint_{\lambda>{\cal C}_{\cal D}}\frac{d\mu}{2\pi i}
B(\mu,\lambda)y(\lambda)\frac{\delta {\cal F}_h}{\delta y(\mu)}
\label{var1}
\\
\frac{\d {\cal F}_h}{\d a_\beta}&=&\oint_{{\cal C}_{\cal D}}\frac{d\mu}{2\pi i}
\frac{\d y(\mu)}{\d a_\beta}\frac{\delta {\cal F}_h}{\delta y(\mu)}
=\oint_{{\cal C}_{a_\beta}}\frac{d\lambda}{2\pi i}
\oint_{\lambda<{\cal C}_{\cal D}}\frac{d\mu}{2\pi i}
B(\mu,\lambda)y(\lambda)\frac{\delta {\cal F}_h}{\delta y(\mu)},
\label{var-a}
\eea
and these expressions differ only by the order of integration.\footnote{We also can substitute
just $y(\lambda)W_h(\lambda)$ for $[y(\lambda)W_h(\lambda)]_-$ when evaluating the residue at
$\lambda=a_\beta$.}

Let us introduce the operator
\be
D_\beta=\frac{\d}{\d a_\beta}-\oint_{{\cal C}_{a_\beta}}\frac{d\lambda}{2\pi i}
y(\lambda)\frac{\d}{\d V(\lambda)}.
\label{da}
\ee
When acting on functionals of $y(\mu)$, it can be presented as
\bea
D_\beta {\cal F}_h[y]&=&
\oint_{{\cal C}_{a_\beta}}\frac{d\lambda}{2\pi i}
\oint_{\lambda<{\cal C}_{\cal D}}\frac{d\mu}{2\pi i}
B(\mu,\lambda)y(\lambda)\frac{\delta {\cal F}_h}{\delta y(\mu)}-
\oint_{{\cal C}_{a_\beta}}\frac{d\lambda}{2\pi i}
\oint_{\lambda>{\cal C}_{\cal D}}\frac{d\mu}{2\pi i}
B(\mu,\lambda)y(\lambda)\frac{\delta {\cal F}_h}{\delta y(\mu)}=
\nonumber
\\
&=&\oint_{{\cal C}_{a_\beta}}\frac{d\lambda}{2\pi i}
y'(\lambda)\frac{\delta {\cal F}_h}{\delta y(\lambda)},
\label{da1}
\eea
and we can schematically present the action of $D_\beta$ as
\be
\hbox{
\begin{picture}(240,30)(10,35)
\thinlines
\curve(85,40, 88.6,50, 95,56.4, 105,60)
\curve(90,40, 92.7,47.5, 97.5,52.3, 105,55)
\curve(95,40, 96.8,45, 100,48.2, 105,50)
%\curve(95.2,38.6, 96.8,35, 100,31.8, 103.4,30.2)
\curve(101,45, 98,48, 95.5,50.5)
\thicklines
\qbezier[8](88.4,48.8)(92.7,47.5)(96.8,45)
\curve(101.4,43.4, 100.4,41.2, 100.2,40,
100.4,38.8, 101.4,36.6, 103.4,35.4, 105,35, 106.6,35.4, 108.6,36.6, 109.6,38.8, 110,40,
109.6,41.2, 108.6,43.4, 106.6,44.6, 105,45, 103.4,44.6)
\put(102,44){\circle{2}}
\put(120,40){\makebox(0,0)[cc]{$-$}}
\end{picture}
\begin{picture}(0,20)(200,35)
\thinlines
\curve(85,40, 88.6,50, 95,56.4, 105,60)
\curve(90,40, 92.7,47.5, 97.5,52.3, 105,55)
\curve(95,40, 96.8,45, 100,48.2, 105,50)
%\curve(95.2,38.6, 96.8,35, 100,31.8, 103.4,30.2)
\curve(88.5,55.5, 92,53, 95,51)
\thicklines
\qbezier[8](88.6,50)(92.7,47.5)(96.8,45)
\curve(80,40, 83.5,52,86.5,55.5)
\curve(88.5,57.5, 93,61.5, 105,65)
\put(87.5,56.5){\circle{2}}
\put(120,40){\makebox(0,0)[cc]{$=$}}
\end{picture}
\begin{picture}(0,20)(150,35)
\thinlines
\curve(85,40, 86,46, 88.6,50)
\curve(95,56.4, 99,59, 105,60)
\curve(95,40, 96.8,45, 100,48.2, 105,50)
%\curve(95.2,38.6, 96.8,35, 100,31.8, 103.4,30.2)
%\curve(88.5,55.5, 92,53, 95,51)
%\qbezier[8](86,58)(93,57)(94,51)
\qbezier[8](88.6,50)(92.7,47.5)(96.8,45)
\thicklines
\curve(90,40, 92.7,47.5, 97.5,52.3, 105,55)
%\curve(80,40, 83.5,52,86.5,55.5)
%\curve(88.5,57.5, 93,61.5, 105,65)
\put(95,50){\circle{2}}
\put(87,54.5){\makebox(0,0)[cc]{${}_{y'(\mu)}$}}
\end{picture}
}
\label{Faa}
\ee
where we let the white dot represent the function $y(\mu)$ and the
fat contour designate the integration over the circle about the branching
point $a_\beta$; apparently, integrations about other branching points commute with this
integration. We can now use representation (\ref{loopmod}), which implies that, actually,
when acting on ${\cal F}_h$, we can put all variations to the outer line to obtain, instead
of (\ref{Faa}),
\be
\hbox{
\begin{picture}(240,40)(10,30)
\put(50,40){\makebox(0,0)[cc]{$D_\beta{\cal F}_h=$}}
\thinlines
\curve(85,40, 88.6,50, 95,56.4, 105,60)
\curve(90,40, 92.7,47.5, 97.5,52.3, 105,55)
\curve(95,40, 96.8,45, 100,48.2, 105,50)
%\curve(95.2,38.6, 96.8,35, 100,31.8, 103.4,30.2)
\qbezier[8](88.4,48.8)(92.7,47.5)(96.8,45)
\thicklines
\curve(101,45, 98,48, 92.5,53.5)
\put(92,54){\circle*{2}}
\put(93,53){\vector(-1,1){0}}
\curve(101.4,43.4, 100.4,41.2, 100.2,40,
100.4,38.8, 101.4,36.6, 103.4,35.4, 105,35, 106.6,35.4, 108.6,36.6, 109.6,38.8, 110,40,
109.6,41.2, 108.6,43.4, 106.6,44.6, 105,45, 103.4,44.6)
\put(102,44){\circle{2}}
\put(120,40){\makebox(0,0)[cc]{$-$}}
\end{picture}
\begin{picture}(0,20)(200,30)
\thinlines
\curve(85,40, 88.6,50, 95,56.4, 105,60)
\curve(90,40, 92.7,47.5, 97.5,52.3, 105,55)
\curve(95,40, 96.8,45, 100,48.2, 105,50)
%\curve(95.2,38.6, 96.8,35, 100,31.8, 103.4,30.2)
\qbezier[8](88.6,50)(92.7,47.5)(96.8,45)
\thicklines
\curve(83.5,58.5, 87.5,57, 91.5,53.5)
\put(91.5,53.5){\circle*{2}}
\put(91,54){\vector(1,-1){0}}
\curve(75,40, 78.5,53,81.5,58.5)
\curve(83.5,60.5, 92,66.5, 105,70)
\put(82.5,59.5){\circle{2}}
\put(120,40){\makebox(0,0)[cc]{$=$}}
\end{picture}
\begin{picture}(0,20)(150,30)
\thinlines
\curve(90,40, 92.7,47.5, 97.5,52.3, 105,55)
\curve(95,40, 96.8,45, 100,48.2, 105,50)
\qbezier[8](88.6,50)(92.7,47.5)(96.8,45)
\thicklines
\curve(85,40, 88.6,50, 90.5,52.5)
\curve(92.5,54.5, 95,56.4, 105,60)
%\put(95,50){\circle{2}}
\put(91.5,53.5){\makebox(0,0)[cc]{$\ast$}}
\end{picture}
}
\label{Faa1}
\ee
where the asterisk now designates the whole combination
$\sum_{h'=1}^{h-1}W_h'(\lambda)W_{h-h'}(\lambda)+W_{h-1}(\lambda,\lambda)$ in the
r.h.s. of (\ref{loopmod}). However, for these considerations to work we must be
able to collapse integration contours to the branching points to produce representation
(\ref{Bertrand}) for the one-loop resolvent. So the above considerations are
only plausible
reasonings, and in order to prove the consistency conditions rigorously we must use other
technique. At present, the only systematic way to prove
these conditions is to use the {\em induction} in the framework of the above
diagrammatic technique. This proof is however too cumbersome to be presented here and will
be published elsewhere.

\section{Calculation in genus one \label{s:genus}}

This is the only place where we integrate the expressions for $W_h(\lambda)$ explicitly in
the spirit of \cite{ACKM} and \cite{Ak96}.

\subsection{Finding the free energy}

\paragraph{Choosing a special basis.}

We now express the action of operator (\ref{1**}) on the monomials $(\mu-\mu_i)$ in terms of
the derivatives w.r.t. $\d/\d V(\mu)$.

Using the above conditions and formula (\ref{A-cycle-F}), we can invert the operator
$\widehat K-2W_0(\lambda)$ when acting on basis monomials $(\lambda-\mu_\alpha)^{-k}$.
We define the basis vectors $\chi_i^{(k)}(\lambda)$ to be
\be
\label{*1*}
\chi_i^{(k)}(\lambda)d\lambda =\oint_{{\cal C}_{\{\mu_j\}}}\frac{d\mu }{2\pi i}\,\frac{d{\cal E}(\lambda ,\mu )}{y(\mu )}
\cdot\frac{1}{(\mu -\mu_i)^k}
\equiv \widehat{d{\cal E}}\left((\lambda -\mu_i)^{-k}\right),\quad i=1,\dots,2n.
\ee
Since the r.h.s. of (\ref{loopmod}) is this operator acting on a collection of monomials, the
general form of the one-point resolvent is
\be
W_h(\lambda)  = \sum_{k=1}^{3h-1}\sum_{i=1}^{2n} A_{i,h}^{(k)} \chi_i^{(k)}(\lambda),
\quad h\ge 1,
\label{Wstr}
\ee
where $A_{\alpha,h}^{(k)}$ are certain functions of $\mu_\alpha$, $a_\beta$, and the
moments $M_i^{(k)}$. As the order of the highest singularity term
$1/\bigl((\lambda-\mu_\alpha)^{3h-1}\ty(\lambda)\bigr)$ in $W_h(\lambda)$
is insensitive to a multi-cut structure, $W_h(\lambda)$ will
depend on at most $2n(3h-2)$ moments, just like the one-cut solution \cite{ACKM}.
Here we need only few first basis functions.

To obtain $\chi_\alpha^{(1)}$, let us start with (\ref{dVy}). It implies
\be
\left.\frac{1}{\lambda-\mu_\alpha}\frac{\d \mu_\alpha}{\d V(\mu)}\right|_{\lambda\to\mu_\alpha}
=2(B(\lambda,\mu)-B(\lambda^*,\mu))\frac1{y(\lambda)}+O(1),
\label{5.1}
\ee
whence, as $\int\frac{d\lambda}{\sqrt{\lambda-\mu_\alpha}}=2\sqrt{\lambda-\mu_\alpha}$,
we have
\be
\left.\frac{\d \mu_\alpha}{\d V(\mu)}\right|_{\lambda\to\mu_\alpha}
=\oint_{{\cal C}_{\mu_\alpha}}\frac1{2\pi i}
\frac{d{\cal E}(\mu,\lambda)}{\lambda-\mu_\alpha}\frac1{y(\lambda)}
=\widehat{d{\cal E}}\left(\frac{1}{\lambda-\mu_\alpha}\right).
\label{5.2}
\ee

It then technically useful to introduce the function
\be
y_\alpha(\mu)\equiv y(\mu)(\mu-\mu_\alpha)^{-1/2},
\label{alpha}
\ee
which, together with all its derivatives in $\mu$, is nonsingular as $\mu\to\mu_\alpha$.
We then apply integration by parts to obtain
\bea
&{}&\oint_{{\cal C}_{\{\mu_j\}}}\frac{d\mu }{2\pi i}\,\frac{d{\cal E}(\lambda ,\mu )}{y(\mu )}
\cdot\frac{1}{(\mu -\mu_\alpha)^2}
=\oint_{{\cal C}_{\mu_\alpha}}\frac{d\mu }{2\pi i}\frac{1}{\mu -\mu_\alpha}
\frac{\d}{\d\mu}\left(\frac{d{\cal E}(\lambda ,\mu )}{y(\mu )}\right)
\nonumber
\\
&{}&\quad=\oint_{{\cal C}_{\mu_\alpha}}\frac{d\mu }{2\pi i}\frac{1}{\mu -\mu_\alpha}
\left(\frac{(B(\lambda ,\mu )-B(\lambda^*,\mu))}{y(\mu )}+
\frac{d{\cal E}(\lambda ,\mu )}{y(\mu )}\left[-\frac{y'_\alpha(\mu)}{y_\alpha(\mu)}-\frac1{2(\mu-\mu_\alpha)}
\right]\right).
\eea
Taking the term with $\mu-\mu_\alpha$ in the square brackets in the r.h.s. to the l.h.s., taking into account
that the term with the Bergmann kernel in the r.h.s. is just $-\d y(\mu)/\d V(\lambda)$, and using (\ref{5.2}),
we obtain
\bea
&{}&\frac32\oint_{{\cal C}_{\mu_\alpha}}\frac{d\mu }{2\pi i}\,\frac{d{\cal E}(\lambda ,\mu )}{y(\mu )}
\cdot\frac{1}{(\mu -\mu_\alpha)^2}
=\oint_{{\cal C}_{\mu_\alpha}}\frac{d\mu }{2\pi i}\frac{1}{\mu -\mu_\alpha}
\left(-\frac{\d y(\mu)}{\d V(\lambda)}\frac{1}{y(\mu )}+
\frac{d{\cal E}(\lambda ,\mu )}{y(\mu )}\left[-\frac{\d}{\d\mu}\log y_\alpha(\mu)\right]\right)
\nonumber
\\
&{}&\quad
=\oint_{{\cal C}_{\mu_\alpha}}\frac{d\mu }{2\pi i}\frac{1}{\mu -\mu_\alpha}
\left(-\frac{\d }{\d V(\lambda)}\log y_{\alpha}(\mu)+\frac{1}{2(\mu-\mu_\alpha)}\right)
-\frac{d\mu_\alpha}{\d V(\lambda)}\left.\frac{\d}{\d\mu}\right|_{\mu=\mu_\alpha}\log y_\alpha(\mu)
\nonumber
\\
&{}&\quad
=\oint_{{\cal C}_{\mu_\alpha}}\frac{d\mu }{2\pi i}\left(-\frac{1}{\mu -\mu_\alpha}
\frac{\d }{\d V(\lambda)}\log y_{\alpha}(\mu)
-\frac{d\mu_\alpha}{\d V(\lambda)}\frac1{(\mu -\mu_\alpha)^2}
\log y_\alpha(\mu)\right)
\nonumber
\\
&{}&\quad
=-\frac{\d }{\d V(\lambda)}\left(\oint_{{\cal C}_{\mu_\alpha}}\frac{d\mu }{2\pi i}\frac{1}{\mu -\mu_\alpha}
\log y_{\alpha}(\mu)\right)
\nonumber
\\
&{}&\quad
=-\frac{\d }{\d V(\lambda)}\left(\log y_{\alpha}(\mu_\alpha)\right)
\label{5.4}
\eea

The last nontrivial action to consider is
\be
\oint_{{\cal C}_{\{\mu_j\}}}\frac{d\mu }{2\pi i}\,\frac{d{\cal E}(\lambda ,\mu )}{y(\mu )}
\cdot\frac{1}{(\mu -a_\beta)^2}=2\frac{B(\lambda,[a_\beta])}{y([a_\beta])}.
\label{5.5}
\ee
On the other hand,
\bea
\frac{\d }{\d V(\lambda)}M(a_\beta)&=&
\frac{\d }{\d V(\lambda)}\oint_{{\cal C}_{a_\beta}}\frac{d\mu }{2\pi i}
\frac{y(\mu )}{\mu-a_\beta}\,\frac1{\ty(\mu)}
\nonumber
\\
&=&
\oint_{{\cal C}_{a_\beta}}\frac{d\mu }{2\pi i}\left(
\frac{-B(\lambda,\mu )}{\mu-a_\beta}\,\frac1{\ty(\mu)}+\frac{M(\mu)}{\mu-a_\beta}
\sum_{\alpha=1}^s\frac1{2(\mu-\mu_\alpha)}\cdot\frac{\d \mu_\alpha}{\d V(\lambda)}\right)
\nonumber
\\
&=&-\frac{B(\lambda,[a_\beta])M(a_\beta)}{y([a_\beta])}+M(a_\beta)
\sum_{\alpha=1}^s\frac1{2(a_\beta-\mu_\alpha)}\cdot\frac{\d \mu_\alpha}{\d V(\lambda)}.
\label{5.6}
\eea

Collecting all the results, we have
\bea
\chi_\alpha^{(1)}(\lambda) &=& \widehat{d{\cal E}}\left(\frac{1}{\lambda-\mu_\alpha}\right)=
\frac{\d \mu_\alpha}{\d V(\lambda)},\quad \alpha=1,\dots,s,
\nonumber\\
\chi_\alpha^{(2)}(\lambda) &=& \widehat{d{\cal E}}\left(\frac{1}{(\lambda-\mu_\alpha)^2}\right)=
-\frac{2}{3}\frac {\d}{\d V(\lambda)} \left(\log M_\alpha^{(1)}\frac{\prod_{\gamma\ne\alpha}^s\sqrt{\mu_\alpha-\mu_\gamma}}
{\prod_{\beta=1}^t\sqrt{\mu_\alpha-a_\beta}}\right),\quad \alpha=1,\dots,s,
\nonumber
\\
\chi_\beta^{(1)}(\lambda) &=& \widehat{d{\cal E}}\left(\frac{1}{\lambda-a_\beta}\right)=0,\quad \beta=1,\dots,t,
\nonumber
\\
\chi_\beta^{(2)}(\lambda) &=& \widehat{d{\cal E}}\left(\frac{1}{(\lambda-a_\beta)^2}\right)=
-2\frac {\d}{\d V(\lambda)}\left(\log M_\beta^{(1)}\prod_{\alpha=1}^s\sqrt{\mu_\alpha-a_\beta}\right),\quad \beta=1,\dots,t.
\label{basis2}
\eea

\paragraph{Integrating $W_1(\lambda)$.}

Now we evaluate the action of $\widehat{d{\cal E}}$ on $W_0(\lambda,\lambda)$ in the
special basis above to obtain the subleading term ${\cal F}_1$ for the free energy.
We first obtain the explicit formula for $W_0(\lambda,\lambda)$ by differentiating expression (\ref{*dE})
w.r.t. $\mu$, subtracting the term $1/(\lambda-\mu)^2$, and taking the limit $\mu\to\lambda$:
\bea
\frac{\d}{\d V(\lambda)} W_0(\lambda)&=&-\frac14\frac{\why''(\lambda)}{\why(\lambda)}
-\frac14\sum_{i=1}^{n-1}H_i(\lambda)\left(\sum_{j=1}^{2n}\frac{1}{\lambda-\mu_j}
\oint_{A_i}d\xi \frac{1}{(\xi-\mu_j)\why(\xi)}\right)
\nonumber
\\
&=& \frac{1}{16}\sum_{i=1}^{2n}\frac{1}{(\lambda-\mu_i)^2}
- \frac{1}{8}\sum_{i,j=1 \atop i<j}^{2n}\frac{1}{\mu_i-\mu_j}\left(\frac{1}{\lambda-\mu_i}-
\frac{1}{\lambda-\mu_j}\right)
\nonumber
\\
&&-\frac14\sum_{i=1}^{n-1}H_i(\lambda)\left(\sum_{j=1}^{2n}\frac{1}{\lambda-\mu_j}
\oint_{A_i}d\xi \frac{1}{(\xi-\mu_j)\why(\xi)}\right).
\label{dW0}
\eea

Turning to the last term containing hyperelliptic integrals, we now prove that
\bea
\sum_{i=1}^{n-1}{ H}_i(\mu_j)\cdot
\oint_{A_i}\frac{d\xi}{(\xi-\mu_j){\why}(\xi)}
&=&\sum_{i=1}^{n-1}\oint_{A_i}\frac{ H_i(\xi)}{(\xi-\mu_j)\why(\xi)}d\xi
\nonumber
\\
&=&2\frac{\d}{\d \mu_j}\log\det\sigma,
\quad j=1,\dots,2n,
\label{main}
\eea
where the matrix $\sigma$ has entries
\be
\label{Q}
\sigma_{i,k}\equiv\oint_{A_i}\frac{\xi^{k-1}}{\why(\xi)}d\xi,
\quad i,k=1,\dots,n-1.
\ee
For the canonical polynomials
$ H_k(\xi)\equiv \sum_{l=1}^{n-1} H_{l,k}\xi^{l-1}$, \ $k=1,\dots,n-1$,
related to the canonically normalized differentials (\ref{canoDV}), i.e.,
such that
$\oint_{A_i}\frac{ H_k(\xi)}{\why(\xi)}d\xi=\delta_{k,i}$, we obviously have
\be
\label{inverse}
\sum_{l=1}^{n-1} \sigma_{i,l} H_{l,k}=\delta_{i,k}\quad\hbox{for}\quad i,k=1,\dots,n-1.
\ee
Therefore, for all $k>0$ such that $l-k-1\ge0$,
\be
\sum_{i=1}^{n-1} { H}_{l,i}\cdot
\oint_{A_i}\frac{\xi^{l-k-1}}{{\why}(\xi)}d\xi=0,,
\ee
and
$$
\sum_{i=1}^{n-1} \oint_{A_i}\frac{{ H}_i(\xi)-{ H}_i(\mu_j)}
{(\xi-\mu_j){\why}(\xi)}d\xi=
\sum_{i=1}^{n-1} \oint_{A_i}\frac{\sum_{l=2}^{n-1}{ H}_{l,i}
\sum_{k=1}^{l-1}\xi^{l-k-1}\mu_j^{k-1}}
{{\why}(\xi)}d\xi=0,
$$
so we obtain the first part of (\ref{main}). To obtain the second part, note
that by virtue of (\ref{inverse}) it is just
$$
\sum_{k=1}^{n-1}\left(2\frac{\d}{\d \mu_j}\sigma_{j,k}\right)\sigma^{-1}_{k,i},
$$
so we come to the second equality in (\ref{main}).

To find $W_1(\lambda)$ it remains to act by $\widehat{d{\cal E}}$ on
(\ref{dW0}) using (\ref{basis2}). This eventually gives
\bea
W_1(\lambda)  &=&\frac{\d}{\d V(\lambda)}\left[
-\frac1{24}\log \prod_{\alpha=1}^s M_\alpha^{(1)}-\frac1{8}\log \prod_{\beta=1}^t M_\beta^{(1)}
-\frac16\log\prod_{1\le\alpha<\gamma\le s} |\mu_\alpha-\mu_\gamma|\right.
\nonumber
\\
&{}&\left.\quad
-\frac16\log\prod_{\alpha=1}^s\prod_{\beta=1}^t |\mu_\alpha-a_\beta|
-\frac 12\log\det\sigma\right]
\label{W1}
\eea
This expression is easy to integrate. Up to a function ${\cal F}(\{a_\beta\})$
depending only on $a_\beta$, we have
\be
\label{F1}
{\cal F}_1=-\frac1{24}\log\left(\prod_{\alpha=1}^{s}M(\mu_\alpha)\cdot\prod_{\beta=1}^{t}M^3(a_\beta)
\cdot\Delta(\mu)^{4}\cdot\prod_{\alpha,\beta}(\mu_\alpha-a_\beta)^4\cdot
(\det_{i,j=1,\dots,n-1}\sigma_{j,i})^{12}\right),
\ee
where $\Delta(\mu)=\prod_{1\leq\alpha<\gamma\leq s}(\mu_\alpha-\mu_\gamma)$ is the
Vandermonde determinant. This is our final answer for the genus-one partition
function. A natural candidate for a function ${\cal F}(\{a_\beta\})$
in this case might be a power of the Vandermonde
determinant composed from $a_\beta$. But, as we demonstrate in the next subsection,
explicitly solving the consistency conditions, we obtain ${\cal F}(\{a_\beta\})=0$ and so
result (\ref{F1}) is {\em exact}.

\subsection{Verifying the consistency conditions}

Using operator $D_\beta$ (\ref{da}), we can present consistency conditions (\ref{match-a})
in the case of ${\cal F}_1$ in the form
\be
D_\beta{\cal F}_1=\oint_{{\cal C}_{a_\beta}}\frac{d\xi}{2\pi i}W_0(\xi,\xi)
\label{comp1}
\ee
with ${\cal F}_1$ from (\ref{F1}) and $W_0(\xi,\xi)$ from (\ref{dW0}). To obtain the answer we
need only to define the action of $D_\beta$ on the moments and on the branching point parameters.
First, from (\ref{resab}), taking the leading term of $B(\lambda,\mu)$ as $\lambda\to\mu_\alpha$,
we immediately obtain
\be
D_\beta\mu_\alpha=0,\quad \alpha=1,\dots, s,\qquad D_\beta a_\gamma=\delta_{\beta,\gamma},\quad \gamma=1,\dots,t.
\label{comp2}
\ee
It is also easy to obtain the action of $D_\beta$ on moments $M_\alpha^{(k)}$ and $M_\gamma^{(k)}$
for $\gamma\ne\beta$. Representing
$$
M_i^{(k)}=\oint_{{\cal C}_{\mu_i}}\frac{d\xi}{2\pi i}\frac{y(\xi)}{(\xi-\mu_i)^k \ty(\xi)},
$$
in the case $\mu_i\ne a_\beta$ we obtain that $D_\beta$ gives zero when acting on all $\mu_\alpha$
in the denominator of the integrand, the contours of integration about $a_\beta$ and $\mu_i$ do not
intersect, so the action of $D_\beta$ on $y(\xi)$ also vanishes, and the only contribution comes
from the explicit variation of just the parameter $a_\beta$, which eventually gives
\be
D_\beta M_i^{(k)}=-\frac12 \frac1{(k-1)!}\frac{\d^{k-1}}{\d \xi^{k-1}}\left.\left(
\frac{M(\xi)}{\xi-a_\beta}\right)\right|_{\xi=\mu_i},\quad \mu_i\ne a_\beta.
\label{comp3}
\ee
The last quantity to calculate is the variation of $M_\beta^{(k)}$. Here, as above, all variations in
$\mu_\alpha$ disappear, but contribution when varying $y(\xi)$ is nonzero because of contour ordering.
Permuting contours of integration as in (\ref{Faa}), we obtain extra derivative in $\xi$, so, denoting by
$\delta_{a_\beta}$ the explicit variation w.r.t. the parameter $a_\beta$, we obtain
$$
D_\beta M_\beta^{(k)}
=\oint_{{\cal C}_{a_\beta}}\frac{d\xi}{2\pi i}y(\xi)\left(\frac{\d}{\d\xi}+\delta_{a_\beta}\right)
\frac{1}{(\xi-a_\beta)^k\ty(\xi)}
=\oint_{{\cal C}_{a_\beta}}\frac{d\xi}{2\pi i}\frac{y(\xi)}{(\xi-a_\beta)^{k-1/2}}\frac{\d}{\d\xi}
\frac{1}{\ty_\beta(\xi)},
$$
where, as in (\ref{alpha}), we let $\ty_\beta(\xi)\equiv \ty(\xi)\sqrt{\xi-a_\beta}$ denote the regular
part of the expression $\ty(\xi)$ in the local coordinates near $a_\beta$. We then eventually obtain
\be
D_\beta M_\beta^{(k)}=\frac{1}{(k-1)!}\frac{\d^{k-1}}{\d\xi^{k-1}}\left.\left(
-\frac12\sum_{\alpha=1}^s\frac{M(\xi)}{\xi-\mu_\alpha}
+\frac12 \sum_{\gamma=1\atop \gamma\ne\beta}^t\frac{M(\xi)}{\xi-a_\gamma}
\right)\right|_{\xi=a_\beta},\quad \beta=1,\dots,t.
\label{comp4}
\ee

Substituting these formulas in (\ref{comp1}) and evaluating the both sides of the equation
using explicit formulas (\ref{F1}) and (\ref{dW0}), we find the
exact coincidence, which means that no additional factors depending exclusively on the hard wall
parameters $a_\beta$ are required to satisfy the consistency conditions.
Answer (\ref{F1}) is therefore exact, and we show below that it also agrees
with the ideology of Bonnet, David, and Eynard~\cite{David}.

\subsection{Genus-one free energy, determinant representation, and singularities
\label{ss:torus}}

We now discuss the result obtained for the genus-one free
energy. Comparing it with the ``old" matrix-model
two-cut solution~\cite{Ak96} we find the coincidence
up to the modular transformation
that permutes $A$- and $B$-periods. This is exactly because
we set all the $A$-periods (\ref{Sfr})
of $dS$ (\ref{dS}) constant under the action of ${\d\over\d V(\lambda)}$,
see (\ref{dVSi}). On the
contrary, if we impose the condition of zero $B$-periods of
$dS$, which corresponds to the equal ``chemical potentials" $\Pi_i$ in different wells of the
potential \cite{david92}, or to additional minimization of free energy
(\ref{variF}) w.r.t. occupation numbers (\ref{oc}),\footnote{This definitely
implies the independence (vanishing) of $B$-periods of $dS$ under the action of $\d/\d V(\lambda)$.}
we come to the modularly transformed basis of cycles.
In this case, the matrix $\sigma^{(A)}_{i,j}$ of
the $A$-periods of ${x^idx\over \why(x)}$
must be replaced by the matrix $\sigma^{(B)}$
of the corresponding $B$-periods~\cite{KMT}.
Certainly, (\ref{F1})
reproduces the answer of \cite{Kos,DST}\footnote{In \cite{Kos},
the determinant term $\det \sigma$ is omitted from the answer.}
for the generic multi-cut solution.

That the only result of interchanging $A$- and
$B$-periods is the interchanging of the corresponding periods in $\sigma$
implies that $e^{{\F}_1}$ is a density, not a function, on the moduli
space of curves. Indeed, the corresponding
determinants are related as $\det \sigma^{(B)}=\det \tau\det \sigma^{(A)}$,
where $\tau_{i,j}=\oint_{B_i}dw_j$ is the period matrix, which is itself related to
${\cal F}_0$ (see (\ref{tau})).
In order to compensate this determinant
$e^{{\F}_1}$ must be transformed under interchanging $A$- and $B$-cycles
with the additional factor $\left(\det\tau_{ij}\right)^{1/2}$, which
is character for a section of determinant bundle of the operator
$\bar\d$ over the moduli space.

It was proposed in \cite{DST} that, in order to match
the proper behavior under modular transformations, the operator $\bar\d$
must act on the twisted bosons on the hyperelliptic curves;
the factor $e^{{\F}_1}$ is then equal to its
determinant. Besides, one also needs to add some corrections due to
the star operators \cite{DST,Moore} that do not contain $\det\sigma$
factors and cannot be restored by the modular invariance
alone. These corrections are however necessary to obtain correct result
(\ref{F1}).

It seems plausible that we can add arbitrary
function of the occupation numbers $S_i$ and $t_0$ to ${\cal F}$
not spoiling the solution to the loop equation. However, if we assume
normalizing conditions (\ref{G}) and (\ref{Z}), which follow from the
condition of the ``locality'' of ${\cal F}$, that is, from the condition
that ${\cal F}$ depends only on local properties of $dS$ near the branching
points, we completely fix this ambiguity. The answer then
becomes singular as $S_i\to 0$, and this is the singularity we are going to discuss now.

The first type of singular behavior occurs when we
shrink a cut with two ``dynamical'' branching points, e.g., bring $\mu_2$ to
$\mu_1$. Setting $\mu_2-\mu_1=\epsilon\to 0$, we first consider what happens to
the determinant term when closing the cycle $A_1$. We have
$$
\lim_{\mu_2\to\mu_1=\mu}\oint_{A_1}\frac{\xi^{j-1}d\xi}{\sqrt{(\xi-\mu_1)(\xi-\mu_2)\cdots(\xi-\mu_{2n})}}
=\frac{\mu^{j-1}}{\sqrt{(\mu-\mu_3)\cdots(\mu-\mu_{2n})}}
$$
and
$$
\lim_{\mu_2\to\mu_1=\mu}\oint_{A_j}\frac{\xi^{j-1}d\xi}{\sqrt{(\xi-\mu_1)(\xi-\mu_2)\cdots(\xi-\mu_{2n})}}
=\oint_{A_j}\frac{\xi^{j-1}d\xi}{(\xi-\mu)\sqrt{(\xi-\mu_3)\cdots(\xi-\mu_{2n})}},\quad j\ne1.
$$
Introducing $\why_{(n-1)}(\xi)=\sqrt{(\xi-\mu_3)\cdots(\xi-\mu_{2n})}$, we then have for the determinant
$$
\lim_{\mu_2\to\mu_1=\mu}\sigma_{(n)}=\frac{1}{\sqrt{(\mu-\mu_3)\cdots(\mu-\mu_{2n})}}\cdot\det
\left(%
\begin{array}{cccc}
  1 & \oint_{A_2}\frac{d\xi}{(\xi-\mu)\why_{(n-1)}(\xi)} & \dots &  \oint_{A_g}\frac{d\xi}{(\xi-\mu)\why_{(n-1)}(\xi)} \\
  \mu &  \oint_{A_2}\frac{\xi d\xi}{(\xi-\mu)\why_{(n-1)}(\xi)} & \dots &
      \oint_{A_g}\frac{\xi d\xi}{(\xi-\mu)\why_{(n-1)}(\xi)} \\
  \vdots & \vdots & \ddots & \vdots \\
  \mu^{n-2} &  \oint_{A_2}\frac{\xi^{n-2}d\xi}{(\xi-\mu)\why_{(n-1)}(\xi)} & \dots &
      \oint_{A_g}\frac{\xi^{n-2}d\xi}{(\xi-\mu)\why_{(n-1)}(\xi)} \\
\end{array}%
\right)
$$
and subtracting now the first column multiplied by $\oint_{A_k}\frac{d\xi}{(\xi-\mu)\why_{(n-1)}(\xi)}$ from the
$k$th column for all $k=2,\dots,g$, we obtain just the determinant of $\sigma_{(n-1)}$ with the prefactor $1/\why_{(n-1)}(\mu)$:
\be
\lim_{\mu_2\to\mu_1=\mu}\det\sigma_{(n)}=\frac{1}{\sqrt{(\mu-\mu_3)\cdots(\mu-\mu_{2n})}}\det\sigma_{(n-1)}.
\label{shr1}
\ee
If we denote $\mu_1-\mu_2=\epsilon$, then in the limit as $\mu_1\to\mu_2=\mu$, $S_1=\epsilon^2\ty_{(n-1)}(\mu)M(\mu)$
and the products of branching points are then combined exactly in a way to give
\be
\lim_{\mu_2\to\mu_1=\mu}{\cal F}_1^{(n)}={\cal F}_1^{(n-1)}-\frac1{12}\log S_1+O(\epsilon),
\ee
where we must substitute in the expression for ${\cal F}_1^{(n-1)}$ the new function ${\overline M}(\xi)=M(\xi)(\xi-\mu)$
with the new zero at $\mu$ obtained by colliding the branching points $\mu_1$ and $\mu_2$.

The singular part of ${\cal F}_1$ as $S_i\to0$ has therefore the structure
\be
({\cal F}_1)_{\hbox{\small sing}}=-\frac1{12}\sum_{i=1}^n \log S_i, \quad S_n\equiv t_0-\sum_{j=1}^{n-1}S_j.
\ee
This is, in fact, a general phenomenon in the matrix model calculations:
at any order of 't~Hooft expansion, the
only potential source for the singular contribution comes from degenerate geometry
of curves, and it is related with the normalization factor in the matrix integral, that is,
with the volume of (the orbit of) the unitary (sub)group.
Indeed, the integral itself is a Taylor series
in $S_i$'s (see formula (4.8) in \cite{KMT}), while the unitary group volume
\cite{versus} in the case of Hermite polynomial distribution
contributes the factor
\be
\int_{-\infty}^{+\infty}\prod_{k=1}^{S_i/\hbar}d\lambda_k \Delta^2(\lambda)e^{-\frac{t_2}{2\hbar}\sum_{k=1}^{S_i/\hbar}\lambda_k^2}
=\left(\frac{\hbar}{t_2}\right)^{\frac{S_i^2}{2\hbar^2}}\prod_{k=1}^{S_i/\hbar}\Gamma (k)\equiv e^{\frac{S_i^2}{2\hbar^2}\log(\hbar/t_2)}
G_2(S_i/\hbar),
\ee
with $G_2(x)$ being the Barnes function \cite{Barns} whose asymptotic expansion as $x\to\infty$
reads~\cite{Barns,d}
\be
\log G_2(x)={x^2\over 2}\log x -{1\over 12}\log x-{3\over 4}x^2 +{1\over 2}x\log 2\pi+\zeta'(-1)+\sum_{h=2}{B_{2h}\over
4h(h-1)}{1\over x^{2h-2}},
\ee
where $B_{2h}$ are Bernoulli coefficients and $\zeta (s)$ is the Riemann
$\zeta$-function. Discarding all polynomial terms, we obtain that the singular
contribution as $S_i\to0$ is \cite{KMT}
\be
\left({\cal F}_0\right)_{\hbox{\small sing}}=\frac{S_i^2}{2}\log S_i,\qquad
\left({\cal F}_1\right)_{\hbox{\small sing}}=-\frac{1}{12}\log S_i,\qquad
\left({\cal F}_h\right)_{\hbox{\small sing}}={B_{2h}\over
4h(h-1)}{1\over S_i^{2h-2}},\quad h\ge 2,
\label{sing1}
\ee
in the case where two dynamical branching points collide.
This is the simplest way to segregate singular contribution; for instance,
direct calculations demonstrate that the answer for
genus-two correction in the one-cut case~\cite{ACKM}
has the prescribed singularity $-\frac1{240}\frac{1}{t_0^{2}}$.\footnote{In the proper normalization in which
$t_0=d^2/16$ for $M(\xi)\equiv 1$.}

In the hard-wall case, we also have
the singularity  appearing when a dynamical point collides with the
hard wall, $\mu_1\to a_1$. In this case, we must calculate the asymptotic expressions for Laguerre,
not Hermite, polynomials as the characteristic expression in this case is
\be
\int_{0}^{+\infty}\prod_{k=1}^{S_i/\hbar}d\lambda_k \Delta^2(\lambda)e^{-\frac{t_1}{\hbar}\sum_{k=1}^{S_i/\hbar}\lambda_k}
=\left(\frac{\hbar}{t_1}\right)^{\frac{S_i^2}{\hbar^2}}\prod_{k=1}^{S_i/\hbar}(\Gamma (k))^2\equiv e^{\frac{S_i^2}{\hbar^2}\log(\hbar/t_1)}
(G_2(S_i/\hbar))^2,
\ee
because the norm $h_k$ for the normalized Laguerre polynomial $\frac{\d^n}{dx^n}(x^ne^{-x})$ is $(n!)^2$.
Then exactly the same considerations yield the corresponding singularities:
\be
\left({\cal F}_0\right)_{\hbox{\small sing}}=S_i^2\log S_i,\qquad
\left({\cal F}_1\right)_{\hbox{\small sing}}=-\frac{1}{6}\log S_i,\qquad
\left({\cal F}_h\right)_{\hbox{\small sing}}={2B_{2h}\over
4h(h-1)}{1\over S_i^{2h-2}},\quad h\ge 2.
\label{sing2}
\ee
Comparing with result (\ref{F1}) using the
same considerations as above (in particular, the behavior of $\det\sigma$ is exactly the same), we find that
\be
\lim_{\mu_1\to a_1=\mu}{\cal F}_1^{(n)}={\cal F}_1^{(n-1)}-\frac1{6}\log S_1+O(\epsilon),
\label{sing3}
\ee
where $S_1\sim \epsilon M(\mu)\ty_{(n-1)}(\mu)$ and the function $M(\xi)$ remains unchanged because
no new zeros or poles
arise when $\mu_1\to a_1$. This demonstrates again the consistency of result (\ref{F1}) because the presence
of additional factors depending on parameters $a_\beta$ in ${\cal F}_1$ would
affect the (nonsingular) part of relation (\ref{sing3}).

\section{Conclusion}

In conclusion, we briefly mention some related papers. Perhaps the first consideration of the
WDVV equations for the differential $dS$ with square-root singularities at the denominator was
done from the matrix-model standpoint by Itoyama and Morozov \cite{IM}. There, it was used to
regularize the matrix-model integral. Worth mentioning is also recent paper \cite{GMR} in
which the CFT technique has been used for finding multiresolvents for the simplest, one-cut zero-potential
matrix model with two hard edges. The asymptotic
expansion for such an integral (the Legendre polynomials) reads
\bea
\log\int_{-a}^a \Delta^2(\lambda)\prod_{k=1}^{t_0/\hbar}d\lambda_k &=&
\frac{t_0^2}{\hbar^2}\log(a/8)+\frac{t_0}{\hbar}\log 2\pi+3\zeta'(-1)-\frac14 \log(t_0/\hbar)
\nonumber
\\
&{}&\quad+\frac{1}{12}\log 2+\sum_{h=2}^\infty\left(\frac{\hbar}{t_0}\right)^{2h-2}\frac{B_{2h}}{h(h-1)}\left(1-\frac{1}{2^{2h}}\right),
\label{Legendre}
\eea
and in particular we see that the third derivative of the leading term vanishes, as follows from (\ref{resgen}).

\section*{Acknowledgments}

The author is grateful to his collaborators B.~Eynard, A.~Marshakov, A.~Mironov, and D.~Vassiliev with whom a part
of results of this paper was obtained. He also acknowledges the useful discussion with M. Bertola, D. Korotkin, and J. Harnad.

This work has been partially financially supported by the
RFBR Grant No.~05-01-00498, Grants of Support for the Scientific
Schools 2052.2003.1, by the Program Mathematical Methods of
Nonlinear Dynamics, by the ANS Grant ``G\'eom\'etrie et Int\'egrabilit\'e en Physique Math\'ematique"
(contract number ANR-05-BLAN-0029-01), and
by the European Community through the FP6
Marie Curie RTN {\em ENIGMA} (Contract number MRTN-CT-2004-5652).

\end{document}